\documentclass[11pt]{article}
\usepackage[DIV13]{typearea}
\usepackage{amsmath} 
\usepackage{amssymb}   
\usepackage{amsfonts}
\usepackage{amscd}
\usepackage{graphicx}
\usepackage[footnotesize,bf]{caption2}
\usepackage{cite}
\usepackage{bbold}
\usepackage{longtable}
\usepackage[center,footnotesize,hang]{subfigure}
\usepackage{url}
\usepackage{appendix}
\usepackage{dcolumn}
\usepackage{bm}

\newcommand{\Eqref}[1]{Eq.\eqref{#1}}

\newcommand{\Tabref}[1]{Table \ref{#1}}

\newcommand{\Secref}[1]{Section \ref{#1}}
\newcommand{\Appref}[1]{Appendix \ref{#1}}

\newcommand{\Rep}[1]{\underline{\mbox{\textbf{#1}}}}
\newcommand{\MoreRep}[2]{\underline{\mbox{\textbf{#1}}} _{\mbox{\textbf{#2}}}}
\newcommand{\Groupname}[2]{$ {#1} _{#2} $}

\newcommand{\VEV}[1]{\langle #1 \rangle}
\newcommand{\La}{\Lambda}

\begin{document}

\begin{titlepage}
\begin{flushright}
SISSA 09/2009/EP
\end{flushright}
\vspace*{5mm}

\begin{center}
{\Large\sffamily\bfseries
\mathversion{bold} The Cabibbo Angle in a Supersymmetric $D_{14}$ Model
\mathversion{normal}} \\[13mm]
{\large
A.~Blum$^{a}$~\footnote{E-mail: \texttt{alexander.blum@mpi-hd.mpg.de}} and
C.~Hagedorn$^{a,b}$~\footnote{E-mail: \texttt{hagedorn@sissa.it}}} 
\\[5mm]
{\small \textit{$^a$ 
Max-Planck-Institut f\"{u}r Kernphysik\\ 
Postfach 10 39 80, 69029 Heidelberg, Germany\\
$^b$ Scuola Internazionale Superiore di Studi Avanzati (SISSA)\\
via Beirut 4, I-34014 Trieste, Italy\\
and\\ INFN, Sezione di Trieste, Italy
}}
\vspace*{1.0cm}
\end{center}
\normalsize
\begin{abstract}
\noindent We construct a supersymmetric model with the flavor symmetry $D_{14}$
in which the CKM matrix element $|V_{ud}|$ can take the value
 $|V_{ud}| =\cos \left( \frac{\pi}{14} \right) \approx 0.97493$
implying that the Cabibbo angle $\theta_C$ is $\sin \left( \theta_C \right) \approx |V_{us}| \approx \sin \left( \frac{\pi}{14} \right)
\approx 0.2225$. These values are very close to those observed in experiments.
 The value of $|V_{ud}|$ ($\theta_{C}$) is based on the fact that different
$Z_{2}$ subgroups of $D_{14}$ are conserved in the up and down quark sector.
In order to achieve this, $D_{14}$ is accompanied by a $Z_{3}$ symmetry. 
The spontaneous breaking of $D_{14}$ is induced by flavons, which are scalar
gauge singlets. The quark mass hierarchy
is partly due to the flavor group $D_{14}$ and partly due to a Froggatt-Nielsen 
symmetry $U(1)_{FN}$ under which only the right-handed quarks transform. The model is
completely natural in the sense that the hierarchies among the quark
masses and mixing angles are generated with the help of symmetries. The 
issue of the vacuum alignment of the flavons is solved 
up to a small number of degeneracies, leaving four different possible values for $|V_{ud}|$.
Out of these, only one of them leads to a phenomenological viable model. 
A study of the $Z_2$ subgroup breaking terms
shows that the results achieved in the symmetry limit 
are only slightly perturbed. At the same time
they allow $|V_{ud}|$ ($\theta_C$) to be well inside the small experimental error bars.
\end{abstract}
\end{titlepage}
\setcounter{footnote}{0}

%%%%%%%%%%%%%%%%%%%%%%%%%%%%%%%%%%
\section{Introduction}
%%%%%%%%%%%%%%%%%%%%%%%%%%%%%%%%%%
\label{sec:intro}

The explanation of the hierarchy among the charged fermion masses and of the peculiar fermion mixings, especially in the lepton sector,
is one of the main issues in the field of model building. The prime candidate for the origin of fermion mass hierarchies and mixing patterns seems to
be a flavor symmetry under which the three generations of Standard Model (SM) particles transform in a certain way.
Unlike the majority of studies which concentrate on the leptonic sector we propose a dihedral group 
\footnote{Dihedral symmetries have already been frequently used as flavor symmetries, see \cite{dihedrals}.}, $D_{14}$, as flavor symmetry
to predict the CKM matrix element $|V_{ud}|$ or equivalently the Cabibbo angle $\theta_C$. The crucial aspect in this model is the
fact that $|V_{ud}|$ is given in terms of group theoretical quantities, like
the index $n$ of the dihedral group $D_{n}$, the index $\rm j$ of the representation $\MoreRep{2}{j}$ under which two of the generations of the
(left-handed) quarks transform and the indices $m_{u,d}$ of the subgroups which remain unbroken in the
up, $Z_2 = \langle \mathrm{B} \mathrm{A}^{m_u} \rangle$, and the down quark sector, $Z_2 = \langle \mathrm{B} \mathrm{A}^{m_d} \rangle$. Thereby, $\rm A$
and $\rm B$ are the two generators of the dihedral group.
The general 
formula for $|V_{ud}|$ is \cite{dntheory,thetaC_lam,thetaC_us}
\begin{equation}
\label{eq:Vud}
|V_{ud}|= \left| \cos \left( \frac{\pi \, (m_{u} - m_{d}) \, \mathrm{j}}{n} \right) \right| \; .
\end{equation}
In particular, the Cabibbo angle neither depends on arbitrarily
tunable numbers, nor is it connected to the quark masses as is the case for the Gatto-Sartori-Tonin (GST)
relation \cite{GST}, $\sin \left( \theta_C \right) \approx |V_{us}| \approx \sqrt{m_d/m_s}$. 
The only dependence arises through the fact that the ordering of the mass
eigenvalues determines which element in the CKM mixing matrix is fixed by the group theoretical
quantities. However, since the hierarchy among the quark masses is also naturally accommodated
in our model, partly by the flavor group $D_{14}$ itself and partly by an additional 
Froggatt-Nielsen (FN) symmetry $U(1)_{FN}$ \cite{FN}, this sort of arbitrariness in the determination of
the Cabibbo angle is avoided. \footnote{This cannot, for example, be avoided in the $A_4$ models \cite{A4alignment}, which successfully
predict tri-bimaximal mixing in the lepton sector, since the hierarchy among the light neutrinos,
which determines the ordering of the columns in the lepton mixing matrix, is very mild. Actually,
a certain fine-tuning is necessary to achieve that the atmospheric mass squared difference is larger than the
solar one.}

In this paper we consider as framework the Minimal Supersymmetric SM (MSSM).
The construction used in our model is in several aspects analogous to the one used 
in \cite{A4alignment} to generate tri-bimaximal mixing in the lepton sector with the help of the group $A_{4}$.
The flavor group is broken at high energies through vacuum expectation values (VEVs) of gauge singlets,
the flavons. The prediction of the actual value of the mixing angle 
originates from the fact that different subgroups of the flavor symmetry are conserved
in different sectors (up and down quark sector) of the theory. The separation of these sectors
can be maintained by an additional cyclic symmetry, which is $Z_3$ in our case. The other crucial
aspect for preserving different subgroups is the achievement of a certain
vacuum alignment. As in \cite{A4alignment}, an appropriate flavon superpotential
can be constructed by introducing a $U(1)_R$ symmetry and adding a specific set of scalar fields, the
driving fields, whose $F$-terms are responsible for aligning the flavon VEVs. As we show, the vacuum can be 
aligned such that in the up quark sector a $Z_2$ symmetry with an even index $m_u$ is preserved, whereas
in the down quark sector the residual $Z_2$ symmetry is generated by $\mathrm{B} \mathrm{A}^{m_d}$
with $m_d$ being an odd integer.
Thus, two different $Z_2$ groups are maintained in the sectors. We can set $m_u=0$ without loss of generality.
However, we are unable to predict the exact value of $m_d$ such that our model leads to four possible
scenarios with four different possible values of $|V_{ud}|$. Out of these scenarios only one, namely
$m_d=1$ or $m_d=13$, results in a phenomenologically viable value of $|V_{ud}|$ (and $\theta_C$)
\begin{equation}
\label{eq:Vud_val}
|V_{ud}|=\cos \left( \frac{\pi}{14} \right) \approx 0.97493 \;\;\; \mbox{and} \;\;\;  
\sin \left( \theta_C \right) \approx |V_{us}| \approx \sin \left( \frac{\pi}{14} \right) \approx 0.2225 \; .
\end{equation}
Similar to \cite{A4alignment}, subleading corrections to masses and mixings arise from higher-dimensional operators.
In general they are at most of relative order $\epsilon \approx \lambda^2 \approx 0.04$, so that \Eqref{eq:Vud_val} holds within 
$\pm 0.04$.

The model presented here surpasses the non-supersymmetric one constructed in \cite{thetaC_us}
in several ways. Since the flavor symmetry is broken only
spontaneously at the electroweak scale in the latter model, 
it contains several copies of the SM Higgs doublet.
In contrast to this, the model which we discuss in the following possesses 
the two MSSM Higgs doublets $h_{u}$ and $h_{d}$, which are neutral under the flavor group, and gauge singlets, the flavons and
driving fields, which transform under flavor. The flavons are responsible for breaking the flavor symmetry. 
As a consequence, none of the problems usually present in models with an extended Higgs
doublet sector, such as too low Higgs masses and large flavor changing neutral currents,
is encountered here. Additionally, the problem of the vacuum alignment, which determines the value of 
the Cabibbo angle, is solved, up to a small number of degeneracies. This is
impossible in the case of a multi-Higgs doublet model due to the large number of quartic couplings.
Only a numerical fit can show that (at least) one set of parameters exists 
which leads to the desired vacuum structure. Finally,
the breaking of the flavor group at high energies is also advantageous, because then 
domain walls generated through this breaking \cite{domainwalls} might well be diluted in an inflationary era.

In the class of models \cite{A4alignment} which extends the flavor group $A_4$, being successful in predicting tri-bimaximal 
mixing for the leptons, to the quark sector one usually observes that
the Cabibbo angle $\theta_{C} \equiv \lambda \approx 0.22$ produced is generically only of the order of 
$\epsilon \approx \lambda^2$ and thus too small by a factor of four to five.
 \footnote{This also happens in a recently proposed model using the flavor group $S_4$ \cite{S4model}.}  This observation might
indicate that it is not possible to treat the Cabibbo angle only as a small perturbation in this class of
models. 

The paper is organized as follows: in \Secref{sec:grouptheory} we repeat the necessary group theory
of $D_{14}$ and the properties of the subgroups relevant here. \Secref{sec:outline} contains an
outline of the model in which the transformation properties of all particles under the flavor group are given.
The quark masses and mixings, in the limit of conserved $Z_2$ subgroups in up and down quark sector,
are presented in \Secref{sec:LO_results}. In \Secref{sec:NLO_results}, 
corrections to the quark mass matrices are studied in detail and the results of \Secref{sec:LO_results} 
are shown to be only slightly changed.
The flavon superpotential is discussed in \Secref{sec:flavons}. 
We summarize our results and give a short outlook in \Secref{sec:summary}. 
Details of the group theory of $D_{14}$ such as Kronecker products and Clebsch Gordan coefficients can be found in \Appref{app:groupdetails}. 
In \Appref{app:flavons_NLO} the corrections to the flavon superpotential and the shifts of the flavon VEVs are given.

%%%%%%%%%%%%%%%%%%%%%%%%%%%%%%%%%%
\mathversion{bold}
\section{Group Theory of $D_{14}$}
\mathversion{normal}
%%%%%%%%%%%%%%%%%%%%%%%%%%%%%%%%%%
\label{sec:grouptheory}

In this section we briefly review the basic features of the dihedral group $D_{14}$.
Its order is 28, and it has four one-dimensional irreducible representations which we denote as
$\MoreRep{1}{i}$, $\rm i=1,...,4$ and six two-dimensional ones called $\MoreRep{2}{j}$,
$\rm j=1,...,6$. All of them are real and the representations $\MoreRep{2}{j}$
with an odd index $\rm j$ are faithful. The group is generated by the two elements
$\rm A$ and $\rm B$ which fulfill the relations \cite{dngrouptheory}
\begin{equation}
\label{eq:genrelations}
\mathrm{A}^{14} =\mathbb{1} \;\;\; , \;\;\; \rm B^2=\mathbb{1} \;\;\; ,
\;\;\; \rm ABA=B \; . 
\end{equation}
The generators $\rm A$ and $\rm B$ of the one-dimensional representations read
\begin{subequations}
\begin{eqnarray}
\MoreRep{1}{1} &\;\;\; : \;\;\;& \rm A=1 \; , \;\; B=1\\ 
\MoreRep{1}{2} &\;\;\; : \;\;\;& \rm A=1 \; , \;\; B=-1\\ 
\MoreRep{1}{3} &\;\;\; : \;\;\;& \rm A=-1 \; , \;\; B=1\\ 
\MoreRep{1}{4} &\;\;\; : \;\;\;& \rm A=-1 \; , \;\; B=-1 \; .
\end{eqnarray}
\end{subequations}
For the representation $\MoreRep{2}{j}$ they are two-by-two matrices of the form
\begin{equation}
\label{eq:generators}
\rm A =\left(\begin{array}{cc} 
                           \mathrm{e}^{\left( \frac{\pi i}{7} \right) \, \mathrm{j}} & 0 \\
                            0 & \mathrm{e}^{-\left( \frac{\pi i}{7} \right)
                              \, \mathrm{j}} 
          \end{array}\right) \; , \; \rm B=\left(\begin{array}{cc} 
                                       0 & 1 \\
                                       1 & 0 
                  \end{array}\right) \; .
\end{equation}
Note that we have chosen $\rm A$ to be complex, although all representations of $D_{14}$ are real.
Due to this, we find for $(a_{1},a_{2})^{T}$ forming the doublet $\MoreRep{2}{j}$
that the combination $(a_{2}^{\star}, a_{1}^{\star})^{T}$ transforms as $\MoreRep{2}{j}$ rather than
$(a_{1}^{\star},a_{2}^{\star})^{T}$. The Kronecker products and Clebsch Gordan coefficients
can be found in \Appref{app:groupdetails} and can also be deduced from the general formulae
given in \cite{kronprods,dntheory}.

Since we derive the value of the element $|V_{ud}|$ (the Cabibbo angle $\theta_C$), through a non-trivial
breaking of $D_{14}$ in the up and down quark sector, we briefly comment on the 
relevant type of $Z_{2}$ subgroups of $D_{14}$. 
These $Z_{2}$ groups are generated by an element
of the form $\mathrm{B} \, \mathrm{A}^{m}$ for $m$ being an integer between 0 and 13. With the help
of \Eqref{eq:genrelations} one easily sees that $(\mathrm{B} \, \mathrm{A}^{m})^{2} 
= \mathrm{B} \, \mathrm{A}^{m} \mathrm{B} \, \mathrm{A}^{m} =  
\mathrm{B} \, \mathrm{A}^{m-1} \mathrm{B} \, \mathrm{A}^{m-1}= ... = \mathrm{B}^{2} = \mathbb{1}$. 
For $m$ being even, singlets transforming as $\MoreRep{1}{3}$ are allowed to have a non-vanishing
VEV, whereas $m$ being odd only allows a non-trivial VEV for singlets which transform as $\MoreRep{1}{4}$
under $D_{14}$. Clearly, all singlets transforming in the trivial representation $\MoreRep{1}{1}$
of $D_{14}$ are allowed to have a non-vanishing VEV. Note that however the fields in the representation
$\MoreRep{1}{2}$ are not allowed a non-vanishing VEV, since $\mathrm{B} \mathrm{A}^m=-1$ for
all possible values of $m$.
In the case of two fields $\varphi_{1,2}$ which form a doublet
$\MoreRep{2}{j}$ a $Z_{2}$ group generated by $\mathrm{B} \, \mathrm{A}^{m}$ is preserved, if
\begin{equation}
\label{eq:VEVstructure}
\left( \begin{array}{c} \langle \varphi_{1} \rangle \\
 \langle \varphi_{2} \rangle \end{array} \right) \propto \left(
\begin{array}{c} \mathrm{e}^{-\frac{\pi i \, \mathrm{j} \, m}{7}}\\ 1
\end{array}
\right) \; .
\end{equation}
In order to see this note that the vector given in \Eqref{eq:VEVstructure} is an eigenvector of
the two-by-two matrix $\mathrm{B} \, \mathrm{A}^{m}$ to the eigenvalue +1.
Due to the fact that 
singlets transforming as $\MoreRep{1}{3}$ can only preserve $Z_{2}$ subgroups generated 
by $\mathrm{B} \, \mathrm{A}^{m}$ with $m$ even and singlets in $\MoreRep{1}{4}$ only those with
$m$ odd, it is possible to ensure that the $Z_2$ subgroup conserved in the up quark is different from
the one in the down quark sector. Note that for this purpose the dihedral group has to have an even index, since
only then the representations $\MoreRep{1}{3,4}$ are present \cite{dntheory}.
So, it is not possible to choose $D_7$ as flavor symmetry, as it has been done in 
\cite{thetaC_lam,thetaC_us},
to predict $\theta_{C}$, if distinct values of $m$ in the up quark and down quark sector are 
supposed to be guaranteed by the choice of representations.
One can check that the subgroup preserved by VEVs of the form given in \Eqref{eq:VEVstructure}
cannot be larger than $Z_2$, if the index $\rm j$ of the representation $\MoreRep{2}{j}$ is odd, i.e. the representation is faithful. 
For an even index $\rm j$ the subgroup is a $D_2$ group generated by the two elements $\rm A^7$
and $\mathrm{B} \mathrm{A}^m$ with $m$ being an integer between $0$ and $6$. \footnote{In general, for fields in representations
$\MoreRep{2}{j}$, whose index $\rm j$ has a greatest common divisor with the group index $n$ larger than one, the preserved subgroup
is larger than a $Z_2$ symmetry. In the case under consideration, namely $n=14$, this statement is equivalent to the statement
that the preserved subgroup is larger than $Z_2$, if the index $\rm j$ of the representation $\MoreRep{2}{j}$ is even. 
We note that there is a mistake in the first version of \cite{dntheory} concerning this aspect.} Obviously, in the case that
only flavons residing in representations $\MoreRep{1}{i}$, $\rm i=1,...,4$, acquire a VEV the conserved 
subgroup is also generally larger than only $Z_2$.

%%%%%%%%%%%%%%%%%%%%%%%%%%%%%%%%%%
\section{Outline of the Model}
%%%%%%%%%%%%%%%%%%%%%%%%%%%%%%%%%%
\label{sec:outline}

In our model the left-handed quarks $Q_{1}$ and $Q_2$ are unified into the $D_{14}$ doublet $\MoreRep{2}{1}$, denoted by $Q_D$, 
while the third generation of left-handed quarks $Q_{3}$, 
the right-handed up-type quark $t^c$, and the right-handed down-type quark $s^c$, transform trivially
under $D_{14}$, i.e. as $\MoreRep{1}{1}$. The remaining two generations of right-handed fields, 
i.e. $c^c$ and $u^c$ in the up quark and $d^c$ and $b^c$ in the down quark sector,
are assigned to the one-dimensional representations $\MoreRep{1}{3}$ and $\MoreRep{1}{4}$.
\footnote{The fact that the transformation properties of the right-handed down quark fields
are permuted compared to those of the right-handed up quark fields is merely due to the desire to
arrive at a down quark mass matrix $\mathcal{M}_{d}$ which has a large $(33)$ 
entry, see \Eqref{eq:Md_LO}. However, since this is just a permutation of the right-handed 
fields it is neither relevant for quark masses nor for mixings.}
The MSSM Higgs doublets $h_u$ and $h_{d}$ do not transform under $D_{14}$. Therefore, we need
to introduce gauge singlets, flavons, to form $D_{14}$-invariant Yukawa couplings. These flavons
transform according to the singlets $\MoreRep{1}{1}$, $\MoreRep{1}{3}$, $\MoreRep{1}{4}$ and the
doublets $\MoreRep{2}{1}$, $\MoreRep{2}{2}$ and $\MoreRep{2}{4}$. All Yukawa operators involving flavon fields 
are non-renormalizable and suppressed by (powers of) the cutoff scale $\Lambda$ which is expected to be of the order
of the scale of grand unification or the Planck scale.
Additionally, we have to introduce a symmetry which allows
us to separate the up and down quark sector. The minimal choice of such a symmetry in
this setup is a $Z_{3}$ group. We assign a trivial $Z_{3}$ charge to left-handed quarks, right-handed
up quarks and to the flavon fields $\psi^u_{1,2}$, $\chi^{u}_{1,2}$, $\xi^{u}_{1,2}$ and $\eta^{u}$, which ought to couple dominantly to up quarks.
The right-handed down quarks transform as $\omega^2$ under $Z_{3}$ with $\omega=\mathrm{e}^{\frac{2 \pi i}{3}}$.
The flavon fields $\psi^{d}_{1,2}$, $\chi^{d}_{1,2}$, $\xi^{d}_{1,2}$, $\eta^d$ and $\sigma$, 
mainly responsible for down quark masses, acquire a phase $\omega$ under $Z_{3}$. 
The MSSM Higgs fields transform trivially also under the $Z_3$ symmetry.
Since the right-handed down quarks have charge $\omega^2$ under $Z_{3}$, whereas $Q_{D}$, $Q_{3}$
 and $h_{d}$ are neutral, the bottom quark does not acquire a mass at the
renormalizable level, unlike the top quark. As a result, the hierarchy between the
top and bottom quark is explained without large $\tan \beta = \langle h_{u} \rangle/\langle h_{d} \rangle$.
The hierarchy between the charm and top quark mass, $m_c/m_t \sim \mathcal{O}(\epsilon^2)$ with $\epsilon \approx \lambda^2 \approx 0.04$,
is naturally accommodated in our model. To achieve the correct ratio between strange and bottom quark mass,
$m_{s}/m_{b} \sim \mathcal{O}(\epsilon)$, we apply the FN mechanism. We add the FN
field $\theta$ to our model which is only charged under  $U(1)_{FN}$. Without loss of generality we can assume
that its charge is -1. Note that we distinguish in our discussion between the FN field $\theta$ and the flavon fields
$\psi^u_{1,2}$, $\chi^{u}_{1,2}$, $\xi^{u}_{1,2}$, $\eta^{u}$, $\psi^{d}_{1,2}$, $\chi^{d}_{1,2}$, $\xi^{d}_{1,2}$, $\eta^d$ and 
$\sigma$ which transform non-trivially under $D_{14} \times Z_3$. If we assign 
a $U(1)_{FN}$ charge $+1$ to the right-handed down-type quark $s^c$, we arrive 
at $m_s/m_b \sim \mathcal{O}(\epsilon)$. Finally, to reproduce the hierarchy
between the first generation and the third one, $m_{u}/m_{t} \sim \mathcal{O}(\epsilon^{4})$ 
and $m_{d}/m_{b} \sim \mathcal{O}(\epsilon^{2})$, also the right-handed quarks,
$u^c$ and $d^c$, have to have a non-vanishing $U(1)_{FN}$ charge.
The transformation properties of the quarks and flavons under $D_{14} \times Z_{3}
\times U(1)_{FN}$ are summarized in \Tabref{tab:particles}.
\begin{table}
\hspace{-0.8in}
\parbox{6in}{
\begin{tabular}{|c||c|c|c|c|c|c|c|c||c||c|c|c|c||c|c|c|c|c|}\hline
Field & $Q_D $ & $Q_{3}$ & $u^c$ & $c^c$ & $t^c$ & $d^c$ & $s^c$ & 
$b^c$ & $h_{u,d}$ & 
$\psi^{u}_{1,2}$ & $\chi_{1,2}^{u}$ & $\xi^{u}_{1,2}$ & $\eta^{u}$ &
$\psi^{d}_{1,2}$ & $\chi_{1,2}^{d}$ & $\xi^{d}_{1,2}$ & $\eta^{d}$ & $\sigma$\\ 
\hline
\Groupname{D}{14} & $\MoreRep{2}{1}$ & $\MoreRep{1}{1}$ & $\MoreRep{1}{4}$ & 
$\MoreRep{1}{3}$ & $\MoreRep{1}{1}$  & $\MoreRep{1}{3}$ & 
$\MoreRep{1}{1}$ & $\MoreRep{1}{4}$ & $\MoreRep{1}{1}$ & $\MoreRep{2}{1}$ & $\MoreRep{2}{2}$ &
$\MoreRep{2}{4}$ & $\MoreRep{1}{3}$ & $\MoreRep{2}{1}$ & $\MoreRep{2}{2}$ &
$\MoreRep{2}{4}$ & $\MoreRep{1}{4}$ & $\MoreRep{1}{1}$\\
\Groupname{Z}{3} & $1$ & $1$ & $1$ & $1$ & $1$ & $\omega^2$ & $\omega^2$ & $\omega^2$
& $1$ & $1$ & $1$ & $1$ & $1$ & $\omega$ & $\omega$ & $\omega$ & $\omega$ & $\omega$\\
$U(1)_{FN}$ & 0 & 0 & 2 & 0 & 0 & 1 & 1 & 0 & 0 & 0 & 0 & 0 & 0 & 0 & 0 & 0 & 0 & 0\\ 
\hline
\end{tabular}}
\begin{center}
\normalsize
\begin{minipage}[t]{15cm}
\caption[Particles of the Model]{Particle content of the model. 
Here we display the transformation properties of fermions and scalars under the flavor group 
$D_{14} \times Z_3 \times U(1)_{FN}$. The symmetry $Z_{3}$ separates
the up and down quark sector.
The left-handed quark
doublets are denoted by $Q_D = (Q_{1},Q_{2})^T$, $Q_{1}=(u,d)^{T}$, $Q_{2}=(c,s)^{T}$, $Q_{3}=(t,b)^{T}$ 
and the right-handed quarks by $u^c$, $c^c$, $t^c$
and $d^c$, $s^c$, $b^c$. The flavon fields indexed by a $u$ give masses to the up quarks
only, at lowest order. Similarly, the fields which carry an index $d$ (including the field $\sigma$)
couple only to down quarks at this order.
We assume the existence of a field $\theta$ which is a gauge singlet
transforming trivially under $D_{14} \times Z_3$.
It is responsible for the 
breaking of the $U(1)_{FN}$ symmetry. Without loss of generality its charge under
$U(1)_{FN}$ can be chosen as $-1$. 
Note that $\omega$ is the third root of unity, i.e. $\omega=\mathrm{e}^{\frac{2 \pi i}{3}}$.
\label{tab:particles}}
\end{minipage}
\end{center}
\end{table}
Given these we can write down the superpotential $w$ which consists of two parts
\begin{equation}
w = w_q + w_f \; .
\end{equation}
$w_q$ contains the Yukawa couplings of the quarks and $w_f$ the flavon superpotential responsible
for the vacuum alignment of the flavons. The mass matrices arising from $w_q$ are discussed in
\Secref{sec:LO_results} and \Secref{sec:NLO_results}, while $w_f$ is studied in \Secref{sec:flavons}.

As already explained in the introduction, the prediction of the CKM matrix element $|V_{ud}|$ or equivalently the
Cabibbo angle $\theta_C$ is based on the fact that the VEVs of the flavons $\{ \psi^{u}_{1,2}, \chi_{1,2}^{u}, \xi^{u}_{1,2}, \eta^{u}\}$
preserve a $Z_2$ subgroup of $D_{14}$ which is generated by the element $\mathrm{B} \mathrm{A}^{m_u}$,
whereas the VEVs of $\{ \psi^{d}_{1,2}, \chi_{1,2}^{d}, \xi^{d}_{1,2}, \eta^{d}, \sigma \}$ coupling dominantly to down quarks
keep a $Z_2$ group originating from the element $\mathrm{B} \mathrm{A}^{m_d}$ conserved with $m_u \neq m_d$.
Due to the fact that $\eta^u$ transforms as $\MoreRep{1}{3}$ and
$\eta^d$ as $\MoreRep{1}{4}$ under $D_{14}$ $m_u$ has to be an even integer between $0$ and $12$ and $m_d$
an odd integer between $1$ and $13$, implying the non-equality of $m_u$ and $m_d$. Since $m_u \neq m_d$, it is also
evident that $D_{14}$ is completely broken in the whole theory. As mentioned, the separation of the two
symmetry-breaking sectors is maintained by the $Z_3$ symmetry. 
However, in terms with more than one flavon in the down and more than two flavons in the up quark
sector the fields $\{ \psi^{u}_{1,2}, \chi_{1,2}^{u}, \xi^{u}_{1,2}, \eta^{u}\}$ couple to down quarks and
$\{ \psi^{d}_{1,2}, \chi_{1,2}^{d}, \xi^{d}_{1,2}, \eta^{d}, \sigma \}$ to up quarks so that this separation of the two
sectors is not rigid anymore. Similarly, non-renormalizable operators in the flavon superpotential mix the two different
sectors inducing shifts in the aligned flavon VEVs. This fact is explained in more detail in \Secref{sec:NLO_results} 
and \Secref{sec:flavons_NLO}.
To elucidate the origin of the prediction of $|V_{ud}|$ ($\theta_C$) we first consider in 
\Secref{sec:LO_results} the mass matrices arising in the case that the two different $Z_2$ subgroups remain unbroken in
up and down quark sector. Then we turn in \Secref{sec:NLO_results} to the discussion of the mass matrix structures 
including the subgroup non-preserving 
corrections from multi-flavon insertions and VEV shifts and show that the results achieved in the limit of unbroken $Z_2$ subgroups in both sectors
still hold, especially the prediction of $|V_{ud}|$ ($\theta_C$) is valid up to $\mathcal{O}(\epsilon)$ corrections.

%%%%%%%%%%%%%%%%%%%%%%%%%%%%%%%%%%
\section{Quark Masses and Mixings in the Subgroup Conserving Case}
%%%%%%%%%%%%%%%%%%%%%%%%%%%%%%%%%%
\label{sec:LO_results}

As mentioned above, all Yukawa terms containing up to two flavons in the up and one flavon in the down quark sector preserve a $Z_2$
group generated by $\mathrm{B} \mathrm{A}^{m_u}$ and by $\mathrm{B} \mathrm{A}^{m_d}$, respectively. In the up quark sector the only
renormalizable coupling generates the top quark mass
\begin{equation}
Q_3 \, t^c \, h_u \; .
\end{equation}
Here and in the following we omit order one couplings in front of the operators. The other elements of the third column and the 
$(32)$ element of the up quark mass matrix $\mathcal{M}_u$ arise at the one-flavon level through the terms
\begin{equation}
\frac{1}{\Lambda} (Q_D \psi^u)  t^c h_u \;\;\; \mbox{and} \;\;\; \frac{1}{\Lambda} Q_3 (c^c \eta^u) h_u \;,
\end{equation}
respectively. We denote with $( \cdots )$ the contraction to a $D_{14}$ invariant.
The elements belonging to the upper $1-2$ subblock of $\mathcal{M}_u$ are generated at the level of two-flavon insertions
\begin{eqnarray}
\label{eq:12subblockMu_symm_1}
&& \frac{\theta^2}{\La^4} (Q_D u^c \chi^u \xi^u ) h_u
+ \frac{\theta^2}{\La^4} \left(Q_D u^c (\xi^u)^2 \right) h_u
+ \frac{\theta^2}{\La^4} (Q_D  \psi^u \eta^u u^c) h_u \; , \\
\label{eq:12subblockMu_symm_2}
&& \frac{1}{\La^2} (Q_D c^c \chi^u \xi^u)  h_u
+ \frac{1}{\La^2} \left(Q_D c^c (\xi^u)^2 \right) h_u
+ \frac{1}{\La^2} (Q_D  \psi^u) (\eta^u c^c) h_u \; .
\end{eqnarray}
Thereby, the $(11)$ and $(21)$ entries stem from the terms in \Eqref{eq:12subblockMu_symm_1}, while the terms in \Eqref{eq:12subblockMu_symm_2} are responsible for
the $(12)$ and $(22)$ elements  of $\mathcal{M}_u$. Also the elements of the third column receive contributions from two-flavon
insertions which, however, can be absorbed into the existing couplings (, if we are in the symmetry preserving limit). Therefore, we do not mention
these terms explicitly here. 
Only the $(31)$ element of $\mathcal{M}_u$ vanishes in the limit of an unbroken $Z_2$ subgroup in the up quark sector, since 
the existence of the residual symmetry forbids a non-zero VEV for a flavon (a combination of flavons)
in the $D_{14}$ representation $\MoreRep{1}{4}$ for even $m_u$.  
The VEVs of the fields 
$\psi^{u}_{1,2}$, $\chi_{1,2}^{u}$ and $\xi^{u}_{1,2}$, which preserve a $Z_2$ symmetry generated by the element $\mathrm{B} \mathrm{A}^{m_u}$, 
are of the form
\begin{equation}
\label{eq:VEVs_LO_doublets_u}
\!\!\!\!\!\!\! \!\!\!\!\left( \begin{array}{c} \VEV{\psi_1^u} \\ \VEV{\psi_2^u} \end{array} \right) = v^u \left( \begin{array}{c} 
\mathrm{e}^{-\frac{\pi i m_u}{7}} \\ 1 \end{array} \right) , \;
\left( \begin{array}{c} \VEV{\chi_1^u} \\ \VEV{\chi_2^u} \end{array} \right) = w^u \, \mathrm{e}^{\frac{\pi i m_u}{7}}  
\left( \begin{array}{c} \mathrm{e}^{-\frac{2 \pi i m_u}{7}}  \\  1 \end{array} \right) , \;
\left( \begin{array}{c} \VEV{\xi_1^u} \\ \VEV{\xi_2^u} \end{array} \right) = z^u \, \mathrm{e}^{\frac{2 \pi i m_u}{7}} 
\left( \begin{array}{c} \mathrm{e}^{-\frac{4 \pi i m_u}{7}} \\  1 \end{array} \right)
\end{equation}
together with $\langle \eta_u \rangle \neq 0$. As can be read off from \Eqref{eq:Vud}, only the difference between $m_u$ and $m_d$ is relevant for $|V_{ud}|$. 
Thus, we set $m_u=0$. Obviously, the conserved $Z_2$ group in the up quark sector is then generated by $\rm B$.
The up quark mass matrix has the generic form
\begin{equation}
\label{eq:Mu_LO}
\mathcal{M}_u = \left(
\begin{array}{ccc}
	- \alpha^u_1 \, t^2 \, \epsilon^2  & \alpha^u_2 \, \epsilon^2 
				& \alpha^u_3 \, \epsilon \\
	\alpha^u_1 \, t^2 \, \epsilon^2 & \alpha^u_2 \, \epsilon^2 & \alpha^u_3 \, \epsilon\\
	0 & \alpha^u_4 \, \epsilon & y_t 
\end{array}
\right) \, \langle h_u \rangle
\end{equation}
in the $Z_2$ symmetry limit. The couplings $\alpha^u_i$ and $y_t$ are in general complex. 
The small expansion parameters $\epsilon$ and $t$ are given by
\begin{equation}
\label{eq:epsilon_u}
\frac{v^u}{\La}, \frac{w^u}{\La}, \frac{z^u}{\La}, \frac{\VEV{\eta^u}}{\La} \sim \epsilon \approx \lambda^2 \approx 0.04 
\;\;\; \mbox{and} \;\;\; \frac{\VEV{\theta}}{\La} = t\; ,
\end{equation}
where we assume that all flavon VEVs are of the same order of magnitude. This is partly justified by the fact that they are correlated by the parameters of the
flavon superpotential, see \Eqref{eq:value_wu_zu}. Additionally, we take $t$ and $\epsilon$ to be real and positive and furthermore assume
\begin{equation}
\label{eq:tapproxepsilon}
t \approx \epsilon \approx \lambda^2 \approx 0.04
\end{equation}
in the following. 

We can discuss the down quark mass matrix $\mathcal{M}_d$ in a similar fashion. Taking into account only terms with one 
flavon we can generate apart from the $(33)$ entry of the matrix the elements of the second column, 
\begin{equation}
\frac{1}{\La} Q_3 (b^c \eta^d) h_d\; , \; \frac{\theta}{\La^2} \, Q_3 \, s^c \sigma h_d\;\;\; 
\mbox{and} \;\;\; \frac{\theta }{\La^2} \, (Q_D \psi^d) \, s^c h_d \; .
\end{equation}
Actually, the first term is responsible for the $(33)$ entry, while the second one leads to a non-vanishing $(32)$ entry and the third one gives the dominant 
contribution to the $(12)$ and $(22)$ elements of $\mathcal{M}_d$. The flavon VEVs preserving the subgroups generated by $\mathrm{B} \mathrm{A}^{m}$
are of the form
\begin{equation}
\label{eq:VEVs_LO_doublets_d} 
\!\!\!\!\!\!\!\!\! 
\left( \begin{array}{c} \VEV{\psi_1^d} \\ \VEV{\psi_2^d} \end{array} \right) = v^d \left( \begin{array}{c} \mathrm{e}^{-\frac{\pi i m}{7}} \\ 1 \end{array} \right) \; , \;\;
\left( \begin{array}{c} \VEV{\chi_1^d} \\ \VEV{\chi_2^d} \end{array} \right) = w^d \mathrm{e}^{\frac{\pi i m}{7}} \left( \begin{array}{c} \mathrm{e}^{-\frac{2 \pi i m}{7}} \\ 1 \end{array} \right) \;\;\; \mbox{and} \;\;\;
\left( \begin{array}{c} \VEV{\xi_1^d} \\ \VEV{\xi_2^d} \end{array} \right) = z^d \mathrm{e}^{\frac{2 \pi i m}{7}} \left( \begin{array}{c} \mathrm{e}^{-\frac{4 \pi i m}{7}} \\ 1 \end{array} \right) 
\end{equation}
with $\VEV{\eta^d}$ and $\VEV{\sigma}$ being non-zero. Since we already set $m_u$ to zero, we omitted the subscript $d$ of the parameter $m$ which has to be
an odd integer ranging between $1$ and $13$. As discussed in \Secref{sec:flavons}, the value of $m$ cannot be uniquely fixed through the superpotential $w_f$. 
The form of the down quark mass matrix is then
\begin{equation}
\label{eq:Md_LO}
\mathcal{M}_d = \left(
\begin{array}{ccc}
	 0 & \alpha^d_1 \, t \, \epsilon  & 0\\
	 0 & \alpha^d_1 \, \mathrm{e} ^{- \pi i m/7} \, t \, \epsilon & 0 \\
	 0 & \alpha^d_2 \, t \, \epsilon & y_b \, \epsilon 
\end{array}
\right) \, \langle h_d \rangle \; .
\end{equation}
Again, the couplings $\alpha^d_i$ and $y_b$ are complex. The expansion parameter $\epsilon$ is given by
\begin{equation}
\label{eq:epsilon_d}
\frac{v^d}{\La}, \frac{w^d}{\La}, \frac{z^d}{\La}, \frac{\VEV{\eta^d}}{\La}, \frac{\VEV{\sigma}}{\La} \sim \epsilon \approx \lambda^2 \approx 0.04 \; .
\end{equation}
The assumption that all VEVs are of the same order of magnitude can also be in this case partly derived from the 
flavon superpotential, see \Eqref{eq:value_wd_zd} and \Eqref{eq:value_x_etad}. Additionally, we assume that $\epsilon$ in \Eqref{eq:epsilon_d} is the same
as in \Eqref{eq:epsilon_u}, i.e. the VEVs of all flavons are expected to be of the same order of magnitude.
\Eqref{eq:tapproxepsilon} thus also holds.

For the quark masses we find 
\begin{subequations}
\label{eq:massratios_LO}
\begin{eqnarray}
m_u^2:m_c^2:m_t^2 &\sim& \epsilon^8: \epsilon^4: 1 \; ,\\
m_d^2:m_s^2:m_b^2 &\sim& 0: \epsilon^2: 1 \; ,\\
m_b^2:m_t^2 &\sim& \epsilon^2:1 \; , 
\end{eqnarray}
\end{subequations}
where the third equation holds for small $\tan \beta$. As one can see, the hierarchy among the up quark masses and the ratio
$m_s/m_b$ are correctly reproduced. The down quark mass vanishes at this level and is generated by $Z_2$ symmetry non-conserving two-flavon insertions, 
see \Eqref{eq:masses_NLO}.
The CKM matrix is of the form
\begin{equation}
\label{eq:VCKM_approx_LO}
|V_{CKM}|  = \left( 
\begin{array}{ccc}
	|\cos (\frac{m \, \pi}{14})| & |\sin (\frac{m \, \pi}{14})| & 0 \\
	|\sin (\frac{m \, \pi}{14})| & |\cos (\frac{m \, \pi}{14})| & 0\\
	0   		             & 0			    & 1
\end{array}
\right) +
\left( 
\begin{array}{ccc}
	0  			& \mathcal{O}(\epsilon^4) 	& \mathcal{O}(\epsilon^2)  \\
	\mathcal{O}(\epsilon^2)	& \mathcal{O}(\epsilon^2)	& \mathcal{O}(\epsilon)    \\
	\mathcal{O}(\epsilon) 	& \mathcal{O}(\epsilon)		& \mathcal{O}(\epsilon^2)
\end{array}
\right) \; .
\end{equation}
The elements $|V_{ud}|$, $|V_{us}|$, $|V_{cd}|$ and $|V_{cs}|$ are determined by the group theoretical parameter $m$.
Since $m$ takes odd integer values between $1$ and $13$ we arrive at four possible scenarios: 
If $m$ takes the value $m=1$ (minimal) or $m=13$ (maximal), we arrive at $|V_{ud}|=\cos (\frac{\pi}{14})
\approx 0.97493$. This value is very close to the central one, $|V_{ud}|_{\rm exp} = 0.97419 ^{+0.00022} _{-0.00022}$, \cite{pdg}. 
For the other three elements of the CKM matrix, also only determined by $m$, we then find 
\begin{equation}
|V_{ud}| \approx |V_{cs}| \approx 0.97493 \;\;\; \mbox{and} \;\;\;
|V_{us}| \approx |V_{cd}| \approx 0.2225 \; ,
\end{equation}
which should be compared with the experimental values \cite{pdg}
\begin{equation}
|V_{cs}|_{\rm exp} = 0.97334 ^{+0.00023} _{-0.00023} \; , \;\;\;
|V_{us}|_{\rm exp} = 0.2257 ^{+0.0010} _{-0.0010} \; , \;\;\;
|V_{cd}|_{\rm exp} = 0.2256 ^{+0.0010} _{-0.0010} \; .
\end{equation}

As the experimental errors are very small, the values predicted for $|V_{ud}|$, $|V_{us}|$, $|V_{cd}|$ and $|V_{cs}|$
are not within the error bars given in \cite{pdg}. However, as we show in \Secref{sec:NLO_results} the terms which break the residual
$Z_2$ subgroups, change the values of the CKM matrix elements $|V_{ud}|$, $|V_{us}|$, $|V_{cd}|$ and $|V_{cs}|$
by order $\epsilon$ so that the results of the model agree with the experimental data. The three other possible values for $|V_{ud}|$ which can arise 
are $\cos (\frac{3 \, \pi}{14}) \approx 0.78183$ for $m=3$ and $m=11$, $\cos (\frac{5 \pi}{14}) \approx 0.43388$ if $m=5$ or $m=9$ and finally $|V_{ud}|$ 
vanishes for $m=7$. Thus, $m$ has to be chosen either minimal or maximal to be in accordance with
the experimental observations. The other values cannot be considered to be reasonable, since we cannot expect that the
corrections coming from symmetry breaking terms change the element $|V_{ud}|$ by more than $\epsilon \approx 0.04$. For this reason,
we set $m=1$ in the following discussion. The CKM matrix elements in the
third row and column are reproduced with the correct order of magnitude, apart from $|V_{ub}|$ which is slightly too small, $\epsilon^2 \approx \lambda^4$
instead of $\lambda^3$, and from $|V_{td}|$ which is slightly too large, $\epsilon \approx \lambda^2$ instead of $\lambda^3$.
The value of $|V_{ub}|$ gets enhanced through the inclusion of $Z_2$ symmetry breaking terms. 
In any case by including $Z_2$ symmetry breaking terms it becomes possible to accommodate all experimental data, 
if some of the Yukawa couplings are slightly enhanced or suppressed. 
$J_{CP}$, the measure of CP violation in the quark sector \cite{Jarlskog}, is of the order $\epsilon^3 \approx \lambda^6$ and thus of the correct order of
magnitude.

Finally, we briefly compare the form of the mass matrices $\mathcal{M}_u$ and $\mathcal{M}_d$ to the general
results we achieved in \cite{dntheory}. According to \cite{dntheory}
 the most general mass matrix arising from the preservation of a $Z_2$ subgroup generated by the element $\mathrm{B} \mathrm{A}^{m_{u,d}}$
for left-handed fields transforming as $\MoreRep{2}{1} + \MoreRep{1}{1}$ and right-handed fields as three singlets
is given by
\begin{equation}
\label{eq:general}
\mathcal{M}_k = \left(
\begin{array}{ccc}
	-A_k & B_k & C_k\\
	A_k \, \mathrm{e}^{- \pi i m_k/7} & B_k \, \mathrm{e}^{- \pi i m_k/7}& C_k \, \mathrm{e}^{- \pi i m_k/7}\\
	0 & D_k & E_k 
\end{array}
\right) \;\;\; \mbox{for} \;\;\; k=u,d \; .
\end{equation}
The parameters $A_k$, $B_k$, $C_k$, $D_k$ and $E_k$ contain Yukawa couplings and VEVs and are in general complex. 
Comparing \Eqref{eq:general} with \Eqref{eq:Mu_LO} shows that $\mathcal{M}_u$ is of this form with $m_u=0$.
The down quark mass matrix $\mathcal{M}_d$, given in \Eqref{eq:Md_LO}, equals the matrix in \Eqref{eq:general},
if $m=m_d$ and the parameters $A_d$ and $C_d$ are set to zero. This happens, since our model only contains 
a restricted number of flavon fields and we do not take into account terms with more than one flavon at this level. 

%%%%%%%%%%%%%%%%%%%%%%%%%%%%%%%%%%
\section{Quark Masses and Mixings including Subgroup-Breaking \\Effects}
%%%%%%%%%%%%%%%%%%%%%%%%%%%%%%%%%%
\label{sec:NLO_results}

In this section we include terms which break the residual $Z_2$ symmetries explicitly. These lead to corrections of the
results shown in \Secref{sec:LO_results}. They are generated by multi-flavon insertions in which flavon fields belonging to the
set $\{ \psi^{d}_{1,2}, \chi^{d}_{1,2}, \xi^{d}_{1,2}, \eta^d, \sigma \}$ give masses to up quarks and flavons belonging to 
$\{ \psi^u_{1,2}, \chi^{u}_{1,2}, \xi^{u}_{1,2}, \eta^{u} \}$ to down quarks. Additionally, non-renormalizable terms in the flavon superpotential
lead to complex shifts in the flavon VEVs in \Eqref{eq:VEVs_LO_doublets_u} and \Eqref{eq:VEVs_LO_doublets_d}. They can be parameterized 
as in \Eqref{eq:VEVshifts} in \Secref{sec:flavons_NLO} (with $m=1$). At the same time,  
$v^d$, $v^u$ and $x$ remain free parameters. 
The corrections to the flavon superpotential are discussed in detail in \Secref{sec:flavons_NLO} and \Appref{app:flavons_NLO}.
The analysis given in these sections shows that the generic size of the shifts is
\begin{equation}
\delta \mathrm{VEV} \sim \mathcal{O} \left( \frac{\mathrm{VEV}}{\La} \right) \, \mathrm{VEV} 
\sim \epsilon \, \mathrm{VEV} \; ,
\end{equation}
if all VEVs are of the order $\epsilon \, \La$, see \Eqref{eq:epsilon_u} and \Eqref{eq:epsilon_d}. 
Thus, the VEV shifts inserted in Yukawa terms with $p$ flavons contribute at the same level as
Yukawa terms containing $p+1$ flavons. 

In the up quark sector we find that the $(11)$ and $(21)$ elements receive $Z_2$ symmetry breaking corrections through the following operators
\begin{eqnarray}
\label{eq:11_21_corr}
&& \frac{\theta^2}{\La^4} \left[ (Q_D u^c \delta\chi^u \xi^u ) + (Q_D u^c \chi^u \delta \xi^u ) \right] h_u 
+ \frac{\theta^2}{\La^4} \left(Q_D u^c \xi^u \delta \xi^u \right) h_u
+ \frac{\theta^2}{\La^4}  (Q_D  \delta \psi^u \eta^u u^c)  h_u\\
+   &&\frac{\theta^2}{\La^5} (Q_D \psi^d \chi^d) (\eta^d u^c) h_u
+ \frac{\theta^2}{\La^5} \left( Q_D u^c (\chi^d)^3 \right)  h_u
+ \frac{\theta^2}{\La^5} \left( Q_D u^c (\psi^d)^2 \xi^d \right) h_u 
+ \frac{\theta^2}{\La^5} \left(Q_D u^c \chi^d (\xi^d)^2  \right) h_u \nonumber\\
+&& \frac{\theta^2}{\La^5} \left( Q_D u^c (\chi^d)^2 \xi^d  \right) h_u
+\frac{\theta^2}{\La^5} (Q_D u^c \chi^d \xi^d)  \sigma h_u
+ \frac{\theta^2}{\La^5} \left( Q_D u^c (\xi^d)^2 \right)  \sigma h_u
+ \frac{\theta^2}{\La^5} (Q_D \psi^d) (\eta^d u^c) \sigma h_u \; . \nonumber
\end{eqnarray}
The notation of, for example, $\delta \chi^u$ indicates that the VEV of the fields $\chi^u_{1,2}$, shifted through the non-renormalizable operators correcting
the flavon superpotential, is used, when calculating the contribution to the up quark mass matrix. Thus, all contributions from the operators in the first line 
of \Eqref{eq:11_21_corr} arise from the fact that the VEVs become shifted. Note that we omitted the operator stemming from the shift of the VEV of $\eta^u$, since
this field only transforms as singlet under $D_{14}$ and thus does not possess any special vacuum structure. (We also do this in the following equations.)
The other operators arise from the insertions
of three down-type flavon fields. There exist similar operators containing three up-type flavons. However, these still preserve the
$Z_2$ symmetry present in the up quark sector at lowest order and therefore can be absorbed into the existing couplings. 
Analogously, we find that the following operators give rise to $Z_2$ symmetry breaking contributions to the $(12)$ and $(22)$ elements so that 
also these are no longer equal
\begin{eqnarray}
&& \frac{1}{\La^2} \left[ (Q_D c^c \delta\chi^u \xi^u ) + (Q_D c^c \chi^u \delta \xi^u ) \right] h_u
+ \frac{1}{\La^2} \left(Q_D c^c \xi^u \delta \xi^u \right) h_u
+ \frac{1}{\La^2} (Q_D  \delta \psi^u) (\eta^u c^c) h_u\\
+ && \frac{1}{\La^3} (Q_D c^c \psi^d \chi^d \eta^d) h_u
+ \frac{1}{\La^3} \left( Q_D c^c (\chi^d)^3 \right) h_u
+ \frac{1}{\La^3} \left( Q_D c^c (\psi^d)^2 \xi^d \right) h_u
+ \frac{1}{\La^3} \left(Q_D c^c \chi^d (\xi^d)^2  \right) h_u \nonumber\\
+&& \frac{1}{\La^3} \left( Q_D c^c (\chi^d)^2 \xi^d  \right) h_u
+\frac{1}{\La^3} (Q_D c^c \chi^d \xi^d)  \sigma h_u
+ \frac{1}{\La^3} \left( Q_D c^c (\xi^d)^2 \right)  \sigma h_u 
+ \frac{1}{\La^3} (Q_D c^c \psi^d \eta^d) \sigma h_u \; . \nonumber
\end{eqnarray}
Again, the operators in the first line are associated to the shifted VEVs. The rest of the operators originates from three-flavon insertions
of down-type flavons. The contributions from the analogous operators with up-type flavons can again be absorbed into the existing couplings.
The dominant contribution to the $(13)$ and $(23)$ element which breaks the residual $Z_2$ symmetry stems from the VEV shift of the 
fields $\psi^u_{1,2}$
\begin{equation}
\frac{1}{\La} \, (Q_D \delta \psi^u) \, t^c h_u \; .
\end{equation}
All other contributions up to three flavons are either $Z_2$ symmetry preserving or breaking, but subdominant.
The $(31)$ element which has to vanish in the symmetry limit is generated through the following three-flavon insertions
of down-type flavons
\begin{eqnarray}
\!\!\!\!\! && \frac{\theta^2 }{\La^5} Q_3 (\eta^d u^c) \sigma^2 h_u
+ \frac{\theta^2 }{\La^5} Q_3 (\eta^d u^c) (\psi^d)^2 h_u
+ \frac{\theta^2 }{\La^5} Q_3 (\eta^d u^c) (\chi^d)^2 h_u
+ \frac{\theta^2 }{\La^5} Q_3 (\eta^d u^c) (\xi^d)^2 h_u\\
\!\!\!\!\!  + && \frac{\theta^2 }{\La^5}  Q_3 (\eta^d u^c) (\eta^d)^2 h_u
+ \frac{\theta^2 }{\La^5} Q_3 (u^c \psi^d \chi^d \xi^d) h_u
+ \frac{\theta^2 }{\La^5} Q_3 \left( u^c \psi^d (\xi^d)^2  \right) h_u \; . \nonumber
\end{eqnarray}
Note that there are symmetry-conserving couplings, i.e. operators with three up-type flavons, of the same order. These, however, 
vanish, if the vacuum alignment in \Eqref{eq:VEVs_LO_doublets_u} is applied.
They can only contribute at the next order, if the VEV shifts are taken into account; however, such effects are subdominant.
The $(32)$ and $(33)$ element of the up quark mass matrix already exist at the lowest order and only receive subdominant contributions
from higher-dimensional operators and VEV shifts.
The up quark mass matrix can thus be cast into the form
\begin{equation}
\label{eq:Mu_NLO}
\mathcal{M}_u = \left(
\begin{array}{ccc}
	t^2 \, (-\alpha^u_1 \, \epsilon^2 + \beta_1^u \, \epsilon^3) & \alpha^u_2 \, \epsilon^2 + \beta_2^u \, \epsilon^3 
				& \alpha^u_3 \, \epsilon + \beta_3^u \, \epsilon^2\\
	\alpha^u_1 \, t^2 \, \epsilon^2 & \alpha^u_2 \, \epsilon^2 & \alpha^u_3 \, \epsilon\\
	\beta^u_4 \, t^2 \, \epsilon^3 & \alpha^u_4 \, \epsilon & y_t 
\end{array}
\right) \, \langle h_u \rangle \; .
\end{equation}
We note that without loss of generality we can define the couplings $\alpha^u_{1,2,3}$ and $\beta^u_{1,2,3}$ in such a way that the corrections stemming from
$Z_2$ subgroup breaking terms only appear in the first row of $\mathcal{M}_u$. Due to this and due to the absorption of subdominant contributions
 the couplings $\alpha^u_i$ only coincide at the leading order with those present in \Eqref{eq:Mu_LO}. 
This also holds for $y_t$. Again, all couplings are in general complex.
The matrix in \Eqref{eq:Mu_NLO} is the most general one arising in our model, i.e. all contributions from terms including more than three flavons
can be absorbed into the couplings $\alpha^u_i$, $\beta^u_i$ and $y_t$.

Similarly, we analyze the $Z_2$ symmetry breaking contributions to the down quark mass matrix $\mathcal{M}_d$.
The $(11)$ and $(21)$ element of $\mathcal{M}_d$ are dominantly generated by $Z_2$ symmetry breaking effects from two-flavon
insertions involving one down- and one up-type flavon. We find five independent operators
\begin{eqnarray}
\label{eq:11_21_Md}
&&\frac{\theta}{\Lambda^3} (Q_D \, d^c \,  \xi^d \chi^u)  h_d
+ \frac{\theta}{\Lambda^3} (Q_D \, d^c \, \chi^d \xi^u) h_d
+ \frac{\theta}{\Lambda^3} (Q_D \, d^c \, \xi^d \xi^u) h_d\\ \nonumber
+ &&\frac{\theta}{\Lambda^3} (Q_D \, \psi^d) (\eta^u d^c) h_d
+ \frac{\theta}{\Lambda^3} (Q_D \, d^c \, \eta^d \psi^u) h_d \; . 
\end{eqnarray}
Since they are $Z_2$ symmetry breaking, the $(11)$ and $(21)$ entries are uncorrelated. We note that $Z_2$ symmetry preserving contributions
can only arise, if operators with more than two flavons are considered. However, these are always subdominant compared to
the operators in \Eqref{eq:11_21_Md}. Similar statements apply to the generation of the $(13)$ and $(23)$ element of $\mathcal{M}_d$.
The dominant ($Z_2$ symmetry breaking) contributions stem from the operators
\begin{eqnarray}
&& \frac{1}{\La^2}  (Q_D \, b^c \xi^d\, \chi^u) h_d
+ \frac{1}{\La^2}  (Q_D \, b^c \, \chi^d \xi^u) h_d
+ \frac{1}{\La^2}  (Q_D \, b^c \, \xi^d \xi^u) h_d\\ \nonumber
+ && \frac{1}{\La^2}  (Q_D \, b^c \, \psi^d \eta^u) h_d
+ \frac{1}{\La^2}  (Q_D  \psi^u) (\eta^d b^c) h_d \; .
\end{eqnarray}
The $(12)$ and $(22)$ elements which are already present at the lowest order are corrected by $Z_2$ symmetry breaking terms
from the VEV shift of the fields $\psi^d_{1,2}$
\begin{equation}
\frac{\theta }{\La^2} \, (Q_D \delta\psi^d) \, s^c h_d
\end{equation}
as well as from two-flavon insertions with one up-type and one down-type flavon
\begin{equation}
\frac{\theta }{\La^3} (Q_D  \chi^d \psi^u ) \, s^c h_d
+ \frac{\theta }{\La^3} (Q_D \psi^d \chi^u) \, s^c h_d
+ \frac{\theta }{\La^3} (Q_D \psi^u) \, \sigma s^c h_d \; .
\end{equation}
The $(31)$ entry, which must vanish in the symmetry limit, is generated dominantly by a single operator
\begin{equation}
\frac{\theta}{\La^3} \, Q_3 (\, \eta^u \, d^c) \, \sigma h_d \; .
\end{equation}
Similarly to the up quark mass matrix, the $(32)$ and $(33)$ elements of $\mathcal{M}_d$ also receive contributions from $Z_2$ symmetry
breaking effects, which can be absorbed into the leading order term.
Eventually, the most general form of the down quark mass matrix $\mathcal{M}_d$ in our model reads 
\begin{equation}
\label{eq_Md_NLO}
\mathcal{M}_d = \left(
\begin{array}{ccc}
	 \beta^d_1 \, t \, \epsilon^2 & t \, (\alpha^d_1 \, \epsilon + \beta^d_4 \, \epsilon^2) & \beta^d_5 \, \epsilon^2\\
	 \beta^d_2 \, t \, \epsilon^2 & \alpha^d_1 \, \mathrm{e} ^{- \pi i/7} \, t \, \epsilon & \beta^d_6 \, \epsilon^2\\
	 \beta^d_3 \, t \, \epsilon^2 & \alpha^d_2 \, t \, \epsilon & y_b \, \epsilon 
\end{array}
\right) \, \langle h_d \rangle \; .
\end{equation}
The parameters $\alpha^d_1$ and $\beta^d_4$ have been defined so that $Z_2$ symmetry breaking
contributions only appear in the $(12)$ element. Note again that all parameters $\alpha^d_i$, $\beta^d_i$ and $y_b$ are complex. Also note that
$\alpha^d_i$ and $y_b$ only coincide at leading order with the corresponding parameters in \Eqref{eq:Md_LO} due to the absorption of
subdominant effects.

Before calculating quark masses and mixings the parameters $\beta^u_4$, $\alpha^u_4$, $y_t$, $\beta^d_3$, $\alpha^d_2$ and $y_b$ in the third row of 
$\mathcal{M}_u$ and $\mathcal{M}_d$ are made real by 
appropriate rephasing of the right-handed quark fields. The resulting quark masses are then (for $t \approx \epsilon$)
\begin{subequations}
\label{eq:masses_NLO}
\begin{eqnarray}
m_u^2 = 2 \vert \alpha_1^u \vert^2 \langle h_u \rangle^2 \epsilon^8 + \mathcal{O}(\epsilon^9) \;\;\;\;\;\;\;\;\;\;\;\;\;\;\;\;
& \; , \;\; &
m_d^2 = \frac{1}{2}  \vert \beta_1^d - \beta_2^d e^{\frac{i \pi}{7}} \vert^2 \langle h_d \rangle^2 \epsilon^6+ \mathcal{O}(\epsilon^7) \; , \\
m_c^2 =  2  \frac{\vert \alpha_3^u \alpha_4^u - y_t \alpha_2^u \vert^2}{y_t^2} \langle h_u \rangle^2 \epsilon^4 + \mathcal{O}(\epsilon^5) 
& \; , \;\; &
m_s^2 =  2 \vert \alpha_1^d \vert^2 \langle h_d \rangle^2 \epsilon^4 + \mathcal{O}(\epsilon^5) \; , \\
m_t^2 = y_t^2 \langle h_u \rangle^2 + \mathcal{O}(\epsilon^2) \;\;\;\;\;\;\;\;\;\;\;\;\; \;\;\;\;\;\;\;\;\;\;\;\;
& \; , \;\; &
m_b^2 = y_b^2 \langle h_d \rangle^2 \epsilon^2 + \mathcal{O}(\epsilon^4) \; .
\end{eqnarray}
\end{subequations}
At the subdominant level thus also the correct order of magnitude of the down quark mass is reproduced.
The CKM matrix elements are given by
\begin{subequations}
\begin{eqnarray}
&& \!\!\! \!\!\! \!\!\!\!\!\! \!\!\! \vert V_{ud} \vert = \cos \left( \frac{\pi}{14} \right) + \mathcal{O}(\epsilon) \, , \,\,
\vert V_{cs} \vert = \cos \left( \frac{\pi}{14} \right) + \mathcal{O}(\epsilon) \, , \\
&& \!\!\! \!\!\! \!\!\!\!\!\! \!\!\! \vert V_{us} \vert = \sin \left( \frac{\pi}{14} \right) + \mathcal{O}(\epsilon) \, , \,\,
\vert V_{cd} \vert = \sin \left( \frac{\pi}{14} \right) + \mathcal{O}(\epsilon) \, , \\
&& \!\!\! \!\!\! \!\!\!\!\!\! \!\!\!\vert V_{cb} \vert = \frac{\epsilon}{\sqrt{2}} \left\vert \frac{\beta_5^d+\beta_6^d}{y_b} - \frac{2 \alpha_3^u}{y_t} \right\vert + \mathcal{O}(\epsilon^2)  \, , \,\, 
\vert V_{ts} \vert = \frac{\epsilon}{\sqrt{2}} \left\vert \frac{\beta_5^d + \beta_6^d e^{\frac{i \pi}{7}}}{y_b} - \frac{\alpha_3^u (1+e^{\frac{i \pi}{7}})}{y_t}  \right\vert + \mathcal{O}(\epsilon^2) \\
&& \!\!\! \!\!\! \!\!\!\!\!\! \!\!\!\vert V_{ub} \vert = \frac{\epsilon}{\sqrt{2}} \left\vert \frac{\beta_5^d - \beta_6^d}{y_b} \right\vert + \mathcal{O}(\epsilon^2) \, , \,\,
\vert V_{td} \vert = \frac{\epsilon}{\sqrt{2}} \left\vert \frac{\beta_5^d - \beta_6^d e^{\frac{i \pi}{7}}}{y_b} - \frac{\alpha_3^u (1-e^{\frac{i \pi}{7}})}{y_t}  \right\vert + \mathcal{O}(\epsilon^2) \, , \,\, \\
&& \!\!\! \!\!\! \!\!\!\!\!\! \!\!\!\vert V_{tb} \vert = 1 + \mathcal{O}(\epsilon^2) \, .
\end{eqnarray}
\end{subequations}
As one can see, $|V_{ud}|$, $|V_{us}|$, $|V_{cd}|$ and $|V_{cs}|$, which are determined by the group theoretical indices of this model,
get all corrected by terms of order $\epsilon$, so that they can be in full accordance with the experimental values \cite{pdg}. The elements of the 
third row and column are still of the same order of magnitude in $\epsilon$ after the inclusion of $Z_2$ subgroup breaking terms, apart from $|V_{ub}|$ which gets 
enhanced by $1/\epsilon$. For this reason, $|V_{td}|$ and $|V_{ub}|$ are both slightly larger in our model, 
$|V_{td}|$, $|V_{ub}| \sim \epsilon \approx \lambda^2$, than the measured values, which are of order $\lambda^3$. However, only a moderate tuning is
necessary in order to also accommodate these values. For the Jarlskog invariant $J_{CP}$ we find
\begin{equation}
J_{CP} = \frac{\epsilon^2}{4 y_b^2 y_t} \, \sin \left( \frac{\pi}{7} \right) \, 
\left(2 \, y_b \, \mathrm{Re}\left((\alpha_3^u)^{\ast} (\beta_5^d - \beta_6^d) \right) - y_t \left( \vert \beta_5^d \vert^2 
- \vert \beta_6^d \vert^2 \right) \right) + \mathcal{O}(\epsilon^3).
\end{equation}
Similar to $|V_{ub}|$ $J_{CP}$ gets enhanced by $1/\epsilon$ compared to the result in the symmetry limit.
Thus, it has to be slightly tuned to match the experimental value, $J_{CP, \rm exp} = \left( 3.05 ^{+0.19} _{-0.20} \right) \times 10^{-5}$, \cite{pdg},
which is around $\epsilon^3 \approx \lambda^6$. However, already the factor $\sin \left( \frac{\pi}{7} \right)/4 \approx 0.11$ 
leads to a certain suppression of $J_{CP}$.

%%%%%%%%%%%%%%%%%%%%%%%%%%%%%%%%%%%%%%%%%%%%%
\section{Flavon Superpotential}
%%%%%%%%%%%%%%%%%%%%%%%%%%%%%%%%%%%%%%%%%%%%%
\label{sec:flavons}

%%%%%%%%%%%%%%%%%%%%%%%%%%%%%%%%%%%%%%%%%%%%%
\subsection{Leading Order}
%%%%%%%%%%%%%%%%%%%%%%%%%%%%%%%%%%%%%%%%%%%%%
\label{sec:LO_flavons}

Turning to the discussion of the flavon superpotential $w_f$ we add - analogously to, for example, \cite{AF2} 
- two additional ingredients. First, we introduce a further $U(1)$ symmetry which is an extension of $R$-parity called $U(1)_{R}$.
Second, a set of so-called driving fields whose $F$-terms account for the vacuum alignment of the flavon fields is added to the model. 
Quarks transform with charge $+1$, flavon fields, $h_{u,d}$ and $\theta$ are neutral and driving fields have a charge $+2$ under $U(1)_R$. 
In this way all terms in the superpotential $w_f$ are linear in the driving fields, 
whereas these fields do not appear in the superpotential $w_q$, responsible for the quark masses. 
Since we expect the flavor symmetry to be broken at high energies around the seesaw scale or the scale of grand unification, soft supersymmetry 
breaking effects will not affect the alignment so that considering only the $F$-terms is justified.
The driving fields, required in order to construct $w_f$, can be found in \Tabref{tab:driving}.
\begin{table}
\begin{center}
\begin{tabular}{|c||c|c|c||c|c|c|}\hline
Field & $\psi^{0u}_{1,2}$ & $\varphi^{0u}_{1,2}$ & $\rho^{0u}_{1,2}$ & 
$\psi^{0d}_{1,2}$ & $\varphi^{0d}_{1,2}$ & $\rho^{0d}_{1,2}$\\ 
\hline
\Groupname{D}{14} & $\MoreRep{2}{1}$ & $\MoreRep{2}{3}$ & $\MoreRep{2}{5}$ & 
$\MoreRep{2}{1}$ & $\MoreRep{2}{3}$  & $\MoreRep{2}{5}$\\
\Groupname{Z}{3} & $1$ & $1$ & $1$ & $\omega$ & $\omega$ & $\omega$\\ 
\hline
\end{tabular}
\end{center}
\begin{center}
\normalsize
\begin{minipage}[t]{12cm}
\caption[Driving Fields of the Model]{Driving fields of the model. 
The transformation properties of the driving fields under the flavor 
symmetry $D_{14} \times Z_{3}$. Similar to the flavons none of the
driving fields is charged under $U(1)_{FN}$. The fields indexed with a $u$ ($d$)
drive the VEVs of the flavons giving  masses dominantly to the up (down) quarks.
Note that all these fields have a $U(1)_R$ charge $+2$.
\label{tab:driving}}
\end{minipage}
\end{center}
\end{table}
The flavon superpotential at the renormalizable level consists of two parts
\begin{equation}
w_{f} = w_{f,u} + w_{f,d}
\end{equation}
where $w_{f,u}$ gives rise to the alignment of the flavons with an index $u$, and $w_{f,d}$ to the alignment of the flavons 
coupling mainly to down quarks. We restrict ourselves to the case of spontaneous CP violation in the flavon sector by taking all parameters in $w_f$ to be real. $w_{f,u}$ reads
\begin{eqnarray}\nonumber
w_{f,u} =& & M_{\psi}^u \left(\psi_1^u \psi_2^{0u}+\psi_2^u \psi_1^{0u}\right) + a_u  \left(\psi_1^u \chi_1^u \varphi_2^{0u}+\psi_2^u \chi_2^u \varphi_1^{0u}\right) 
+ b_u \left(\psi_1^u \chi_2^u \psi_1^{0u}+\psi_2^u \chi_1^u \psi_2^{0u}\right) \\
&+& c_u \left(\psi_1^u \xi_2^u \varphi_1^{0u}+\psi_2^u \xi_1^u \varphi_2^{0u}\right) 
+ d_u \eta^u \left(\xi_1^u \varphi_1^{0u} + \xi_2^u \varphi_2^{0u}\right) + e_u \left(\psi_1^u \xi_1^u \rho_2^{0u} + \psi_2^u \xi_2^u \rho_1^{0u}\right) \nonumber\\
&+& f_u \eta^u \left(\chi_1^u \rho_1^{0u}+\chi_2^u \rho_2^{0u}\right). 
\end{eqnarray}
The conditions for the vacuum alignment are given by the $F$-terms
\begin{subequations}
\begin{eqnarray}
\frac{\partial w_{f,u}}{\partial \psi_1^{0u}} &=& M_{\psi}^u \psi_2^u + b_u \psi_1^u \chi_2^u = 0 \; ,
\\ 
\frac{\partial w_{f,u}}{\partial \psi_2^{0u}} &=& M_{\psi}^u \psi_1^u + b_u \psi_2^u \chi_1^u = 0 \; ,
\\ 
\frac{\partial w_{f,u}}{\partial \varphi_1^{0u}} &=& a_u \psi_2^u \chi_2^u+c_u \psi_1^u \xi_2^u+d_u \eta^u \xi_1^u = 0 \; ,\\
\frac{\partial w_{f,u}}{\partial \varphi_2^{0u}} &=& a_u \psi_1^u \chi_1^u+c_u \psi_2^u \xi_1^u+d_u \eta^u \xi_2^u = 0 \; ,\\ 
\frac{\partial w_{f,u}}{\partial \rho_1^{0u}} &=& e_u \psi_2^u \xi_2^u+f_u \eta^u \chi_1^u = 0 \; ,
\\ 
\frac{\partial w_{f,u}}{\partial \rho_2^{0u}} &=& e_u \psi_1^u \xi_1^u+f_u \eta^u \chi_2^u = 0 \; .
\end{eqnarray}
\end{subequations}
If we assume that none of the parameters in the superpotential vanishes and $\psi_1^u$ acquires a non-zero VEV, we arrive at
\begin{equation}
\label{eq:VEVsupsector}
\!\!\!\!\!\!\! \!\!\!\!\left( \begin{array}{c} \VEV{\psi_1^u} \\ \VEV{\psi_2^u} \end{array} \right) = v^u \left( \begin{array}{c} 
\mathrm{e}^{-\frac{\pi i m_u}{7}} \\ 1 \end{array} \right) , \;
\left( \begin{array}{c} \VEV{\chi_1^u} \\ \VEV{\chi_2^u} \end{array} \right) = w^u \, \mathrm{e}^{\frac{\pi i m_u}{7}}  
\left( \begin{array}{c} \mathrm{e}^{-\frac{2 \pi i m_u}{7}}  \\  1 \end{array} \right) , \;
\left( \begin{array}{c} \VEV{\xi_1^u} \\ \VEV{\xi_2^u} \end{array} \right) = z^u \, \mathrm{e}^{\frac{2 \pi i m_u}{7}} 
\left( \begin{array}{c} \mathrm{e}^{-\frac{4 \pi i m_u}{7}} \\  1 \end{array} \right)
\end{equation}
with
\begin{equation}
\label{eq:value_wu_zu}
w^u  =  -\frac{M_{\psi}^u}{b_u} \; , \;\; 
z^u  =  \frac{w^u}{2  d_u e_u} \left( c_u f_u \pm \sqrt{4 a_u d_u e_u f_u+ (c_u f_u)^2} \right)
\;\;\; \mbox{and} \;\;\; \VEV{\eta^{u}}=- \frac{e_u}{f_u} \frac{v^u z^u}{w^u} \, \mathrm{e}^{-\frac{4 \pi i m_u}{7}} 
\end{equation}
as unique solution. The flavon VEVs are aligned and their alignment only depends on the parameter $m_u$ 
which is an even integer between $0$ and $12$ (see \Secref{sec:grouptheory}). 
Thus, all vacua conserve a $Z_2$ subgroup of $D_{14}$ generated by the 
element $\mathrm{B} \, 
\mathrm{A}^{m_u}$. Since only the difference between $m_u$ and $m_d$ is relevant for the prediction of the CKM matrix element $|V_{ud}|$,
we set $m_u=0$, as it has been done in \Secref{sec:LO_results}, when we study quark masses and mixings. The size of the flavon VEVs
is partly determined by the parameters in $w_{f,u}$ and partly by the free parameter $v^u$. However, it is reasonable to
assume that all VEVs are of the same order of magnitude $\epsilon \Lambda$, as done in \Secref{sec:LO_results} and
\Secref{sec:NLO_results}. Choosing the parameters in
$w_{f,u}$ appropriately, we can make all VEVs in \Eqref{eq:VEVsupsector} and \Eqref{eq:value_wu_zu} positive for $m_u=0$.

Analogously, the flavon superpotential which drives the vacuum alignment of the fields $\psi^d_{1,2}$,
$\chi^{d}_{1,2}$, $\xi^{d}_{1,2}$, $\eta^d$ and $\sigma$ is given by
\begin{eqnarray}\nonumber
w_{f,d} = & & m_{\psi}^d \sigma \left(\psi_1^d \psi_2^{0d}+\psi_2^d \psi_1^{0d}\right) + a_d  \left(\psi_1^d \chi_1^d \varphi_2^{0d}+\psi_2^d \chi_2^d \varphi_1^{0d}\right) 
+ b_d \left(\psi_1^d \chi_2^d \psi_1^{0d}+\psi_2^d \chi_1^d \psi_2^{0d}\right) \\
&+& c_d \left(\psi_1^d \xi_2^d \varphi_1^{0d}+\psi_2^d \xi_1^d \varphi_2^{0d}\right) 
+ d_d \eta^d \left(\xi_1^d \varphi_1^{0d} - \xi_2^d \varphi_2^{0d}\right) + e_d \left(\psi_1^d \xi_1^d \rho_2^{0d} + \psi_2^d \xi_2^d \rho_1^{0d}\right) \nonumber\\
&+& f_d \eta^d \left(\chi_1^d \rho_1^{0d}-\chi_2^d \rho_2^{0d}\right). 
\end{eqnarray}
Setting the $F$-terms of the driving fields $\psi^{0d}_{1,2}$, $\varphi^{0d}_{1,2}$ and $\rho^{0d}_{1,2}$
to zero we find 
\begin{subequations}
\begin{eqnarray}
\frac{\partial w_{f,d}}{\partial \psi_1^{0d}} &=& m_{\psi}^d \sigma \psi_2^d + b_d \psi_1^d \chi_2^d = 0 \; ,
\\ 
\frac{\partial w_{f,d}}{\partial \psi_2^{0d}} &=& m_{\psi}^d \sigma \psi_1^d + b_d \psi_2^d \chi_1^d = 0 \; ,
\\ 
\frac{\partial w_{f,d}}{\partial \varphi_1^{0d}} &=& a_d \psi_2^d \chi_2^d+c_d \psi_1^d \xi_2^d+d_d \eta^d \xi_1^d = 0 \; ,
\\ 
\frac{\partial w_{f,d}}{\partial \varphi_2^{0d}} &=& a_d \psi_1^d \chi_1^d+c_d \psi_2^d \xi_1^d-d_d \eta^d \xi_2^d = 0 \; ,
\\ 
\frac{\partial w_{f,d}}{\partial \rho_1^{0d}} &=& e_d \psi_2^d \xi_2^d+f_d \eta^d \chi_1^d = 0 \; ,
\\ 
\frac{\partial w_{f,d}}{\partial \rho_2^{0d}} &=& e_d \psi_1^d \xi_1^d-f_d \eta^d \chi_2^d = 0 \; .
\end{eqnarray}
\end{subequations}
These equations lead to the same VEV structure as shown in \Eqref{eq:VEVsupsector}, if we assume that again none of the parameters 
in the flavon superpotential vanishes and the two fields $\psi_1^d$ 
and $\sigma$ get a non-vanishing VEV. Thus, $\VEV{\psi_{1,2}^d}$, 
$\VEV{\chi_{1,2}^d}$ and $ \VEV{\xi_{1,2}^d}$ have the same form
as $\VEV{\psi_{1,2}^u}$, 
$\VEV{\chi_{1,2}^u}$ and $ \VEV{\xi_{1,2}^u}$ with obvious replacements $\{v^u, w^u, z^u \} \rightarrow \{v^d, w^d, z^d \}
$, $m_u \rightarrow m_d$ and $m_d$ being an odd integer. $w^d$ and $z^d$ are given by
\begin{equation}
\label{eq:value_wd_zd}
w^d= -\frac{m_{\psi}^d x}{b_d} \;\;\; \mbox{and} \;\;\; 
z^d= \frac{w^d}{2 d_d e_d} \left( c_d f_d \pm \sqrt{4 a_d d_d e_d f_d+(c_d f_d)^2} \right) 
\end{equation}
and the VEVs of the two singlets $\sigma$ and $\eta^d$ read
\begin{equation}
\label{eq:value_x_etad}
\VEV{\sigma}= x \;\;\; \mbox{and} \;\;\; 
\VEV{\eta^d}  =  \frac{e_d}{f_d} \frac{v^d z^d}{w^d} \, \mathrm{e}^{-\frac{4 \pi i m_d}{7}}\; .
\end{equation}
Similar to the parameter $v^u$ the VEVs of $\psi_{2}^d$ and $\sigma$, $v^d$ and $x$, are undetermined, so that not all flavon
VEVs have to be of similar size. 
Note further that the two possible signs appearing in \Eqref{eq:value_wu_zu} and  \Eqref{eq:value_wd_zd} are uncorrelated.
We can choose the parameters such that $v^d$, $w^d$, $z^d$ and $x$ are positive.
The parameter $m_d$ is an odd integer in the range
$\{1,...,13 \}$. Similar to $m_u$ being even, $m_d$ is required to be odd by the transformation property of the flavon
$\eta^d$ under $D_{14}$. Especially, $m_d$ is different from $m_u$ so that we preserve
different $Z_2$ subgroups in both sectors. As a consequence, the derived mixing angle is non-trivial. However,
we cannot uniquely fix the parameter $m_d$ and thus the mixing angle 
by the vacuum alignment deduced from $w_f$. As discussed in \Secref{sec:LO_results}, we are left with
 a small number (four) of different possible values for $|V_{ud}|$.
Due to the different subgroups preserved in up and down quark sector $D_{14}$ is eventually completely broken in the whole theory.
As we set $m_u$ already to zero, we omit the index of the parameter
$m_d$ from now on also in the discussion of the superpotential.

We end with a few remarks about the VEVs of the
driving fields, the absence of a $\mu$-term and the mass spectrum of the gauge singlets transforming under $D_{14}$.
The VEVs of the driving fields are determined by the $F$-terms of the flavon fields. If we plug in the
solutions for the VEVs of the flavons found in \Eqref{eq:VEVsupsector}, \Eqref{eq:value_wu_zu} and \Eqref{eq:value_wd_zd} and take into account the
constraints that none of the parameters in $w_f$ should vanish and also not the parameters $v^d$, $v^u$
and $x$, we arrive at the result that the VEVs of all driving fields have to vanish at the minimum unless the parameters of the potential
fulfill a specific relation. The term $\mu h_u h_d$ is forbidden by the $U(1)_R$ symmetry. It cannot be generated through terms including
one driving field, $h_u$ and $h_d$ and an appropriate number of flavon fields (to make it invariant under the symmetry $D_{14} \times Z_3$),
since the driving fields cannot acquire non-vanishing VEVs. Thus, the $\mu$-term has to originate from another mechanism, see also \cite{AF2}.
In the spectrum of the flavon and driving fields we find massless modes in the supersymmetric limit. These are expected to become massive, if soft supersymmetry
breaking masses are included into the potential. Possible flat directions present in the potential in the supersymmetric limit are also expected
to be lifted through soft supersymmetry breaking terms.

%%%%%%%%%%%%%%%%%%%%%%%%%%%%%%%%%%%%%%%%%%
\subsection{Corrections to the Leading Order}
%%%%%%%%%%%%%%%%%%%%%%%%%%%%%%%%%%%%%%%%%%
\label{sec:flavons_NLO}

In the flavon superpotential, terms containing three flavons and one driving field lead to corrections of the vacuum alignment achieved through the 
superpotential $w_f$, i.e.
they induce (small) shifts in the VEVs of the flavons.
Such terms are suppressed by the cutoff scale $\La$. Due to the $Z_{3}$ symmetry two types of three-flavon combinations can couple to a driving field with an index $u$, 
namely either all three flavons also carry an index $u$ or all three of them belong to the set $\{ \psi_{1,2}^d, \chi_{1,2}^d, \xi_{1,2}^d, \eta^d, \sigma \}$. 
If the driving field has an index $d$, two of the three flavons have to be down-type flavons, while the third one necessarily has to carry an index $u$. 
These corrections to the flavon superpotential can be written as
\begin{equation}
\Delta w_{f} = \Delta w_{f,u} + \Delta w_{f,d}
\end{equation}
where the terms of $\Delta w_{f,u}$ ($\Delta w_{f,d}$) are responsible for the shifts in the VEVs of the flavons
uncharged (charged) under the $Z_{3}$ symmetry. The exact form of the terms is given in 
\Appref{app:flavons_NLO}. We choose the following convention for the shifts of the VEVs
\begin{eqnarray}
\label{eq:VEVshifts}
&& \VEV{\psi_2^u}  =  v^u +\delta v^u \; , \;\; 
\VEV{\chi_i^u}  =  w^u +\delta w_i^u \; , \;\;
\VEV{\xi_i^u}   =  z^u +\delta z_i^u \; , \;\;
\VEV{\eta^u}  =  -\frac{e_u}{f_u} \frac{v^u z^u}{w^u} + \delta \eta^u \\ \nonumber
&& \VEV{\psi_2^d}  =  v^d +\delta v^d \; , \;\; 
\VEV{\chi_1^d}  =  \mathrm{e}^{-\frac{\pi i m}{7}} \left(w^d +\delta w_1^d\right) \; , \;\;
\VEV{\chi_2^d}  = \mathrm{e}^{\frac{\pi i m}{7}} \left(w^d + \delta w_2^d\right) \; , \;\; \\ \nonumber
&& \VEV{\xi_1^d}  =  \mathrm{e}^{- \frac{2 \pi i m}{7}}\left(z^d+\delta z_1^d\right) \; , \;\; 
\VEV{\xi_2^d} =  \mathrm{e}^{\frac{2 \pi i m}{7}}\left(z^d+\delta z_2^d\right) 
\;\;\; \mbox{and} \;\;\;
\VEV{\eta^d} =  \mathrm{e}^{-\frac{4 \pi i m}{7}} \left( \frac{e_d}{f_d} \frac{v^d z^d}{w^d} + \delta \eta^d \right) 
\; ,
\end{eqnarray}
while
\begin{equation}
\label{eq:VEVsfree}
\VEV{\psi_1^u}= v^u \; , \;\; \VEV{\psi_1^d} =  v^d \, \mathrm{e}^{-\frac{\pi i m}{7}}
\;\;\; \mbox{and} \;\;\; \VEV{\sigma} = x 
\end{equation}
remain as free parameters. As can be read off from the equations given in \Appref{app:flavons_NLO} $v^u$, $v^d$ and $x$ are also not fixed by the corrections 
to the superpotential. We do not fix the parameter $m$ in \Eqref{eq:VEVshifts}, although we showed in \Secref{sec:LO_results} 
that only $m=1$ and $m=13$ lead to a phenomenologically viable model. This is done, because the complexity of the calculation of the 
shifts does not depend on the 
actual value of $m$. ($m$ still has to be an odd integer.) One finds that also the inclusion of the corrections
to the flavon superpotential does not fix the value of $m$.
The detailed calculations given in \Appref{app:flavons_NLO} show that the generic size of the shifts is 
\begin{equation}
\delta \mathrm{VEV} \sim \mathcal{O} \left( \frac{\mathrm{VEV}}{\La} \right) \, \mathrm{VEV} 
\sim \epsilon \, \mathrm{VEV}
\end{equation}
for all VEVs being of the order $\epsilon \, \La$. The shifts are expected to be in general complex, without having a particular phase.

%%%%%%%%%%%%%%%%%%%%%%%%%%%%%%%%%%%%%%%%%%%%%%%
\section{Summary and Outlook}
%%%%%%%%%%%%%%%%%%%%%%%%%%%%%%%%%%%%%%%%%%%%%%%
\label{sec:summary}

We presented an extension of the MSSM in which the value of the CKM matrix element $|V_{ud}|$ or equivalently the
Cabibbo angle $\theta_C$ is fixed by group theoretical quantities of the flavor symmetry $D_{14}$, up to 
the choice among four different possible values. The determination of $|V_{ud}|$ originates from
the fact that residual $Z_2$ symmetries of $D_{14}$ exist in the up and down quark sector. We have shown that
these can be maintained by the vacuum alignment resulting from a properly constructed flavon superpotential. Furthermore, it is 
ensured through the choice of flavon representations that the $Z_2$ symmetries of the up and down quark sector do not coincide 
so that the quark mixing cannot be
trivial. It turns out that the vacua of $Z_2$ symmetries generated by $\mathrm{B} \mathrm{A}^m$ with $m$ being either
even or odd are degenerate so that we arrive at the mentioned four possible values for $|V_{ud}|$. Out of these
only one is phenomenologically viable, namely $|V_{ud}|=\cos ( \frac{\pi}{14} ) \approx 0.97493$.
The CKM matrix elements $|V_{us}|$, $|V_{cd}|$ and $|V_{cs}|$ are as well predicted to be $|V_{us}| \approx |V_{cd}| \approx 0.2225$
and $|V_{cs}| \approx |V_{ud}| \approx 0.97493$. For the other elements we find the following orders of magnitude in 
$\epsilon \approx \lambda^2$ (including $Z_2$ subgroup breaking effects): $|V_{cb}|$, $|V_{ts}|$, $|V_{ub}|$, $|V_{td}| \sim \epsilon \approx \lambda^2$
and $|V_{tb}| = 1 + \mathcal{O}(\epsilon^2)=  1 + \mathcal{O}(\lambda^4)$.
Thus, $|V_{td}|$ and $|V_{ub}|$ turn out to be slightly too large. The same is true for $J_{CP}$ which is of the order of 
$\epsilon^2 \approx \lambda^4$ instead of $\lambda^6$. However, it only requires a moderate tuning of the parameters of the model to accommodate
the experimentally measured values. All quark masses are appropriately reproduced. The large top quark mass results from the 
fact that the top quark is the only fermion acquiring a mass at the renormalizable level. Since the bottom quark
mass stems from an operator involving one flavon, the correct ratio $m_b/m_t \sim \epsilon$ is produced
without large $\tan \beta$. The hierarchy $m_u : m_c : m_t \sim \epsilon^4 : \epsilon^2 : 1$ in the up quark 
sector is accommodated in the $Z_2$ subgroup conserving limit. Thereby, the suppression of the up quark mass
is (partly) due to the non-vanishing FN charge of the right-handed up quark. The correct order of magnitude of the strange quark mass
can as well be achieved through the FN mechanism. The down quark mass which vanishes at the lowest order is generated
by operators with two-flavon insertions. Also its correct size is guaranteed by the FN mechanism.
The main problem which cannot be solved in this model is the fact that the parameter $m_{(d)}$ - and therefore also $|V_{ud}|$ - is not
uniquely fixed, but can take a certain number of different values. We presume that a new type of mechanism for
the vacuum alignment is necessary which also fixes the (absolute) phase of the VEVs of the flavons so that the parameter
$m_{(d)}$ is determined. One possibility
might arise in models with extra dimensions. For a recent discussion of the breaking of a flavor symmetry
with extra dimensions see \cite{breakingGfinEDs}.

As a next step, it is interesting to discuss the extension of our model to the leptonic sector.
In the literature models with the dihedral flavor group $D_{3}$ ($\cong S_3$) \cite{D3_for_leptons_preserved}
or $D_4$ \cite{D4_for_leptons_preserved} can found which also use the fact that different
subgroups of the flavor symmetry are conserved in the charged lepton and neutrino (Dirac
and right-handed Majorana neutrino) sector to predict the leptonic mixing angle $\theta_{23}$ to be maximal and $\theta_{13}$ to be zero. 
These models are non-supersymmetric and contain Higgs doublets transforming non-trivially under the flavor group
in their original form. However, recently variants of \cite{D4_for_leptons_preserved} have been discussed
whose framework is the MSSM and in which only gauge singlets break the flavor group spontaneously at high energies
\cite{D4extensions}. A possibility to combine such a variant with the model presented here
by using a (possibly larger) dihedral group is worth studying.

As has been discussed in \cite{dntheory}, the assignment $\Rep{2} + \Rep{1}$ for the left-handed
and $\Rep{1}+\Rep{1}+\Rep{1}$ for the right-handed fields is not the only possible one in order to predict
one element of the mixing matrix in terms of group theoretical quantities only. Alternatively, we can consider a model 
in which both, left- and right-handed fields, are assigned to $\Rep{2}+\Rep{1}$. Such an assignment usually emerges
when we consider grand unified theories (GUTs), e.g. in $SU(5)$ where the left- and right-handed up quarks both reside in
the representation $\Rep{10}$. \footnote{The case in which all three generations transform as singlets is not very appealing,
since the up quark sector would be merely determined by an abelian flavor symmetry rather than
a non-abelian one.} However, the following problem might occur: the product $\Rep{2} \times \Rep{2}$ contains an
invariant of the dihedral group, if left- and right-handed fields transform as the same doublet. The group theoretical
reason is the fact that all two-dimensional representations of dihedral groups are real. The existence of the invariant leads to
a degenerate mass spectrum among the first two generations, e.g. in an $SU(5)$ GUT to the prediction that up quark
and charm quark mass are degenerate. One possibility to circumvent this difficulty might be to resort
to a double-valued dihedral group. Such a group additionally possesses pseudo-real
(two-dimensional) representations. One of their properties is that the product of a 
representation with itself contains the invariant/trivial representation $\MoreRep{1}{1}$ in its
anti-symmetric part. In an $SU(5)$ model one can then use the fact that the contribution of a Higgs field in the GUT representation $\Rep{5}$
to the up quark mass matrix leads to a symmetric mass matrix, in order to avoid the 
invariant coupling. However, it is still not obvious whether the mass
hierarchy among the up quarks can be generated (through the FN mechanism)
without tuning the parameters. Even in non-unified models in which the two-dimensional
representations under which left- and right-handed fields transform do not have to be equivalent,
it might not be obvious that the fermion mass hierarchy can be appropriately accommodated (with an additional
FN symmetry). 

Finally, further interesting aspects to analyze are the anomaly conditions holding for the flavor symmetry $D_{14}$
which in general lead to additional constraints \cite{anomalies} as well as the origin of 
such a flavor symmetry, see for instance \cite{origin}.

%%%%%%%%%%%%%%%%%%%%%%%%%%%%%%%%%%%%%%%%%%%%%%%%
\subsection*{Acknowledgements}
%%%%%%%%%%%%%%%%%%%%%%%%%%%%%%%%%%%%%%%%%%%%%%%%

We would like to thank Lorenzo Calibbi, Ferruccio Feruglio and Andrea Romanino for discussions. 
A.B. acknowledges support from the Studienstiftung des Deutschen Volkes.
C.H. was partly supported by the ``Sonderforschungsbereich'' TR27.

\newpage
\appendix

%%%%%%%%%%%%%%%%%%%%%%%%%%%%%%%%%%%%%%%%%%%%%%%
\section{Kronecker Products and Clebsch Gordan Coefficients}
%%%%%%%%%%%%%%%%%%%%%%%%%%%%%%%%%%%%%%%%%%%%%%%
\label{app:groupdetails}

Here we list the explicit form of the Kronecker products as well as the Clebsch Gordan coefficients.
More general results for dihedral groups with an arbitrary index $n$ can be found in 
\cite{kronprods,dntheory}.

%%%%%%%%%%%%%%%%%%%%%%%%%%%%%%%%%%%%%%%%%%%%%%
\subsection{Kronecker Products}
%%%%%%%%%%%%%%%%%%%%%%%%%%%%%%%%%%%%%%%%%%%%%%

The products $\MoreRep{1}{i} \times \MoreRep{1}{j}$ are
\begin{equation}\nonumber
\MoreRep{1}{i} \times \MoreRep{1}{i}= \MoreRep{1}{1} \; , \;\; 
\MoreRep{1}{1} \times \MoreRep{1}{i}= \MoreRep{1}{i} \;\; \mbox{for} \;\; \rm i=1,...,4 
\; , \;\;
\MoreRep{1}{2} \times \MoreRep{1}{3}= \MoreRep{1}{4} \; , \;\;
\MoreRep{1}{2} \times \MoreRep{1}{4}= \MoreRep{1}{3} \;\; \mbox{and} \;\;
\MoreRep{1}{3} \times \MoreRep{1}{4}= \MoreRep{1}{2} \; .
\end{equation}
For $\MoreRep{1}{i} \times \MoreRep{2}{j}$ we find
\begin{equation}\nonumber
\MoreRep{1}{1,2} \times \MoreRep{2}{j}= \MoreRep{2}{j} \;\;\; \mbox{and} \;\;\;
\MoreRep{1}{3,4} \times \MoreRep{2}{j}= \MoreRep{2}{7-j} \;\;\; \mbox{for all} \;\;\; \rm j \; .
\end{equation}
The products of $\MoreRep{2}{i} \times \MoreRep{2}{i}$ decompose into
\begin{equation}\nonumber
\left[ \MoreRep{2}{i} \times \MoreRep{2}{i} \right] = \MoreRep{1}{1} + \MoreRep{2}{j}
\;\;\; \mbox{and} \;\;\;
\left\{ \MoreRep{2}{i} \times \MoreRep{2}{i} \right\} = \MoreRep{1}{2}
\end{equation}
where the index $\rm j$ equals $\rm j=2i$ for $\rm i \leq 3$ and 
$\mathrm{j}= 14 - 2\rm i$ holds for $\rm i \geq 4$. $\left[ \nu \times \nu \right]$ denotes thereby 
the symmetric part of the product $\nu \times \nu$, while $\left\{ \nu \times \nu \right\}$ is the anti-symmetric 
one. For the mixed products $\MoreRep{2}{i} \times \MoreRep{2}{j}$ with $\rm i \neq j$
two structures are possible either
\begin{equation}\nonumber
\MoreRep{2}{i} \times \MoreRep{2}{j} = \MoreRep{2}{k} + \MoreRep{2}{l}
\end{equation}
with $\rm k=|i-j|$ and $\rm l$ being $\rm i+j$ for $\rm i+j \leq 6$ and $14 - (\rm i+j)$ for $\rm i+j \geq 8$.
For $\rm i+j=7$ we find instead
\begin{equation}\nonumber
\MoreRep{2}{i} \times \MoreRep{2}{j} = \MoreRep{1}{3} + \MoreRep{1}{4} + \MoreRep{2}{k}
\end{equation}
where $\rm k$ is again $\rm |i-j|$.

%%%%%%%%%%%%%%%%%%%%%%%%%%%%%%%%%%%%%%%%%%%%%%
\subsection{Clebsch Gordan Coefficients}
%%%%%%%%%%%%%%%%%%%%%%%%%%%%%%%%%%%%%%%%%%%%%%

For $s_i \sim \MoreRep{1}{i}$ and $(a_1,a_2)^{T} \sim \MoreRep{2}{j}$ we find
\begin{equation}\nonumber
\left( \begin{array}{c} s_1 a_1 \\ s_1 a_2
\end{array} \right) \sim \MoreRep{2}{j} \;\; , \;\;\;
\left( \begin{array}{c} s_2 a_1 \\ -s_2 a_2
\end{array} \right) \sim \MoreRep{2}{j} \;\; , \;\;\;
\left( \begin{array}{c} s_3 a_2 \\ s_3 a_1
\end{array} \right) \sim \MoreRep{2}{7-j} \;\;\; \mbox{and} \;\;\;
\left( \begin{array}{c} s_4 a_2 \\ -s_4 a_1
\end{array} \right) \sim \MoreRep{2}{7-j} \;\; .
\end{equation}
The Clebsch Gordan coefficients of the product of $(a_{1},a_{2})^{T}$, $(b_{1},b_{2})^{T}$ 
$\sim \MoreRep{2}{i}$ read
\begin{equation}\nonumber
a_{1} b_{2} + a_{2} b_{1} \sim \MoreRep{1}{1} \; , \;\;
 a_{1} b_{2} - a_{2} b_{1} \sim \MoreRep{1}{2} \; , 
\;\;\;  \left( \begin{array}{c}
	a_{1} b_{1}\\
	a_{2} b_{2}
	\end{array}
	\right) \sim \MoreRep{2}{j} \;\;\; \mbox{or} \;\;\;
 \left( \begin{array}{c}
	a_{2} b_{2}\\
	a_{1} b_{1}
	\end{array}
	\right) \sim \MoreRep{2}{j}
\end{equation}
depending on whether $\rm j= 2 i$ as it is for $\rm i \leq 3$
or $\mathrm{j}= 14 - 2 \rm i$ which holds if $\rm i \geq 4$.
For the two doublets $(a_{1}, a_{2})^{T} \sim \MoreRep{2}{i}$ and
$(b_{1}, b_{2})^{T} \sim \MoreRep{2}{j}$ we find for 
$\rm i + j \neq 7$
\begin{eqnarray}\nonumber
&& \left( \begin{array}{c}
	a_{1} b_{2}\\
	a_{2} b_{1}
	\end{array}
	\right) \sim \MoreRep{2}{k} \;\;\; (\rm k=i-j) \;\;\; \mbox{or} \;\;\;
 \left( \begin{array}{c}
	a_{2} b_{1}\\
	a_{1} b_{2}
	\end{array}
	\right) \sim \MoreRep{2}{k} \;\;\; (\rm k=j-i)
\\ \nonumber
&& \left( \begin{array}{c}
	a_{1} b_{1}\\
	a_{2} b_{2}
	\end{array}
	\right) \sim \MoreRep{2}{l} \;\;\;  (\rm l=i+j) \;\;\;\;\; \mbox{or} \;\;\;
 \left( \begin{array}{c}
	a_{2} b_{2}\\
	a_{1} b_{1}
	\end{array}
	\right) \sim \MoreRep{2}{l} \;\;\; (\rm l=14 -(i+j))
\end{eqnarray}
If $\rm i+j=7$ holds the covariants read
\begin{equation}\nonumber
a_{1} b_{1} + a_{2} b_{2} \sim \MoreRep{1}{3} \; , \;\;
 a_{1} b_{1} - a_{2} b_{2} \sim \MoreRep{1}{4} \; , 
\;\;\; \left( \begin{array}{c}
	a_{1} b_{2}\\
	a_{2} b_{1}
	\end{array}
	\right) \sim \MoreRep{2}{k} \;\;\; \mbox{or} \;\;\;
 \left( \begin{array}{c}
	a_{2} b_{1}\\
	a_{1} b_{2}
	\end{array}
	\right) \sim \MoreRep{2}{k} \; .
\end{equation}
Again, the first case is relevant for $\rm k=i-j$, while the second form for
$\rm k=j-i$.

%%%%%%%%%%%%%%%%%%%%%%%%%%%%%%%%%%%%%%%%%%%%%%%
\section{Corrections to the Flavon Superpotential}
%%%%%%%%%%%%%%%%%%%%%%%%%%%%%%%%%%%%%%%%%%%%%%%
\label{app:flavons_NLO}

In this appendix we discuss the form of the VEV shifts induced by the corrections of the flavon
superpotential. These corrections can be written as
\begin{equation}\nonumber
\Delta w_{f} = \Delta w_{f,u} + \Delta w_{f,d} \; .
\end{equation}
We can parameterize the shifted VEVs as shown in \Eqref{eq:VEVshifts}. 
$v^d$, $v^u$ and $x$ remain unchanged, since they are free parameters. As mentioned, since the complexity of the calculation is not increased, if 
$m$ is not fixed, it is kept as parameter in the VEVs.
For the actual calculation of the shifts we choose a plus sign in $z^u$ and $z^d$ in front of the square root,
see \Eqref{eq:value_wu_zu} and \Eqref{eq:value_wd_zd}.
The corrections to the flavon superpotential, which induce shifts in the VEVs of the fields 
with an index $u$, are given by
\begin{equation}
\Delta w_{f,u} = \frac{1}{\Lambda} \left( \sum_{k=1}^{16} r_k^u I_k^{R,u} + \sum_{k=1}^{11} s_k^u I_k^{S,u} + \sum_{k=1}^{12} t_k^u I_k^{T,u} \right).
\end{equation}
The invariants $I_k^{R,u}$ read
\begin{equation}
\begin{array}{ll}
I_1^{R,u} = \sigma^2 \left(\psi_1^d \psi_2^{0u}+ \psi_2^d \psi_1^{0u}\right) &
I_9^{R,u} = \left(\psi_1^d \psi_2^{0u} + \psi_2^d \psi_1^{0u}\right) \left(\eta^d\right)^2 \\
I_2^{R,u} = \sigma \left(\psi_1^d \chi_2^d \psi_1^{0u} + \psi_2^d \chi_1^d \psi_2^{0u}\right) &
I_{10}^{R,u} = \left(\psi_1^u \psi_2^{0u} + \psi_2^u \psi_1^{0u}\right) \left(\eta^u\right)^2 \\
I_3^{R,u} = \left(\psi_1^d \psi_2^{0u} + \psi_2^d \psi_1^{0u}\right)\left(\psi_1^d \psi_2^d\right) &
I_{11}^{R,u} = \left(\psi_1^d \chi_1^d \xi_2^d \psi_1^{0u} + \psi_2^d \chi_2^d \xi_1^d \psi_2^{0u}\right) \\
I_4^{R,u} = \left(\psi_1^u \psi_2^{0u} + \psi_2^u \psi_1^{0u}\right)\left(\psi_1^u \psi_2^u\right) &
I_{12}^{R,u} = \left(\psi_1^u \chi_1^u \xi_2^u \psi_1^{0u} + \psi_2^u \chi_2^u \xi_1^u \psi_2^{0u}\right) \\
I_5^{R,u} = \left(\psi_1^d \psi_2^{0u} + \psi_2^d \psi_1^{0u}\right) \left(\chi_1^d \chi_2^d\right) &
I_{13}^{R,u} = \eta^d \left(\chi_1^d \xi_1^d \psi_1^{0u} - \chi_2^d \xi_2^d \psi_2^{0u}\right) \\
I_6^{R,u} = \left(\psi_1^u \psi_2^{0u} + \psi_2^u \psi_1^{0u}\right) \left(\chi_1^u \chi_2^u\right) &
I_{14}^{R,u} = \eta^u \left(\chi_1^u \xi_1^u \psi_1^{0u} + \chi_2^u \xi_2^u \psi_2^{0u}\right) \\
I_7^{R,u} = \left(\psi_1^d \psi_2^{0u} + \psi_2^d \psi_1^{0u}\right) \left(\xi_1^d \xi_2^d\right) &
I_{15}^{R,u} = \eta^d \left(\left(\xi_1^d\right)^2 \psi_2^{0u} - \left(\xi_2^d\right)^2 \psi_1^{0u}\right) \\
I_8^{R,u} = \left(\psi_1^u \psi_2^{0u} + \psi_2^u \psi_1^{0u}\right) \left(\xi_1^u \xi_2^u\right) &
I_{16}^{R,u} = \eta^u \left(\left(\xi_1^u\right)^2 \psi_2^{0u} + \left(\xi_2^u\right)^2 \psi_1^{0u}\right) \; .
\end{array}
\end{equation}
For $I_k^{S,u}$ we find
\begin{equation}
\begin{array}{ll}
I_1^{S,u} = \sigma \left(\psi_1^d \chi_1^d \varphi_2^{0u} + \psi_2^d \chi_2^d \varphi_1^{0u}\right) & 
I_7^{S,u} = \left(\psi_1^u \chi_2^u \xi_1^u \varphi_2^{0u} + \psi_2^u \chi_1^u \xi_2^u \varphi_1^{0u}\right) \\
I_2^{S,u} = \sigma \left(\psi_1^d \xi_2^d \varphi_1^{0u} + \psi_2^d \xi_1^d \varphi_2^{0u}\right) &
I_8^{S,u} = \left(\left(\chi_1^d\right)^2 \psi_2^d \varphi_2^{0u} + \left(\chi_2^d\right)^2 \psi_1^d \varphi_1^{0u}\right) \\
I_3^{S,u} = \sigma \eta^d \left(\xi_1^d \varphi_1^{0u} -\xi_2^d \varphi_2^{0u}\right) &
I_9^{S,u} = \left(\left(\chi_1^u\right)^2 \psi_2^u \varphi_2^{0u} + \left(\chi_2^u\right)^2 \psi_1^u \varphi_1^{0u}\right) \\
I_4^{S,u} = \left(\left(\psi_1^d\right)^3 \varphi_2^{0u} + \left(\psi_2^d\right)^3 \varphi_1^{0u}\right) &
I_{10}^{S,u} = \eta^d \left(\left(\chi_1^d\right)^2 \varphi_1^{0u} - \left(\chi_2^d\right)^2 \varphi_2^{0u} \right) \\
I_5^{S,u} = \left(\left(\psi_1^u\right)^3 \varphi_2^{0u} + \left(\psi_2^u\right)^3 \varphi_1^{0u}\right) &
I_{11}^{S,u} = \eta^u \left(\left(\chi_1^u\right)^2 \varphi_1^{0u} + \left(\chi_2^u\right)^2 \varphi_2^{0u} \right) \\
I_6^{S,u} = \left(\psi_1^d \chi_2^d \xi_1^d \varphi_2^{0u} + \psi_2^d \chi_1^d \xi_2^d \varphi_1^{0u}\right) &
\end{array}
\end{equation}
and for $I_k^{T,u}$
\begin{equation}
\begin{array}{ll}
I_1^{T,u} = \sigma \left(\psi_1^d \xi_1^d \rho_2^{0u} + \psi_2^d \xi_2^d \rho_1^{0u}\right) &
I_7^{T,u} = \left(\chi_1^d \xi_1^d \psi_2^d \rho_2^{0u} + \chi_2^d \xi_2^d \psi_1^d \rho_1^{0u}\right) \\
I_2^{T,u} = \sigma \eta^d \left(\chi_1^d \rho_1^{0u} - \chi_2^d \rho_2^{0u}\right) &
I_8^{T,u} = \left(\chi_1^u \xi_1^u \psi_2^u \rho_2^{0u} + \chi_2^u \xi_2^u \psi_1^u \rho_1^{0u}\right) \\
I_3^{T,u} = \eta^d \left(\left(\psi_1^d\right)^2 \rho_1^{0u} - \left(\psi_2^d\right)^2 \rho_2^{0u}\right) &
I_9^{T,u} = \left(\left(\xi_1^d\right)^2 \psi_1^d \rho_1^{0u} + \left(\xi_2^d\right)^2 \psi_2^d \rho_2^{0u}\right) \\
I_4^{T,u} = \eta^u \left(\left(\psi_1^u\right)^2 \rho_1^{0u} + \left(\psi_2^u\right)^2 \rho_2^{0u}\right) &
I_{10}^{T,u} = \left(\left(\xi_1^u\right)^2 \psi_1^u \rho_1^{0u} + \left(\xi_2^u\right)^2 \psi_2^u \rho_2^{0u}\right) \\
I_5^{T,u} = \left(\psi_1^d \left(\chi_1^d\right)^2 \rho_2^{0u} + \psi_2^d \left(\chi_2^d\right)^2 \rho_1^{0u}\right) &
I_{11}^{T,u} = \eta^d \left( \chi_2^d \xi_1^d \rho_1^{0u} - \chi_1^d \xi_2^d \rho_2^{0u}\right) \\
I_6^{T,u} = \left(\psi_1^u \left(\chi_1^u\right)^2 \rho_2^{0u} + \psi_2^u \left(\chi_2^u\right)^2 \rho_1^{0u}\right) &
I_{12}^{T,u} = \eta^u \left( \chi_2^u \xi_1^u \rho_1^{0u} + \chi_1^u \xi_2^u \rho_2^{0u}\right) \; .
\end{array}
\end{equation}
The shifts in the VEVs of the set of fields $\{ \psi^{d}_{1,2}, \chi_{1,2}^{d}, \xi^{d}_{1,2}, \eta^{d}, \sigma \}$ 
originate from non-renormalizable terms which are of the form
\begin{equation}
\Delta w_{f,d} = \frac{1}{\Lambda} \left( \sum_{k=1}^{21} r_k^d I_k^{R,d} + \sum_{k=1}^{14} s_k^d I_k^{S,d} + \sum_{k=1}^{16} t_k^d I_k^{T,d} \right) \; .
\end{equation}
The invariants $I_k^{R,d}$ are the following
\begin{equation}
\begin{array}{ll}
I_1^{R,d} = \sigma^2 \left(\psi_1^u \psi_2^{0d}+ \psi_2^u \psi_1^{0d}\right) &
I_{12}^{R,d} = \left(\psi_1^d \psi_2^{0d} - \psi_2^d \psi_1^{0d}\right) \eta^u \eta^d \\
I_2^{R,d} = \sigma \left(\psi_1^d \chi_2^u \psi_1^{0d} + \psi_2^d \chi_1^u \psi_2^{0d}\right) &
I_{13}^{R,d} = \left(\psi_1^u \psi_2^{0d} + \psi_2^u \psi_1^{0d}\right) \left(\eta^d\right)^2 \\
I_3^{R,d} = \sigma \left(\psi_1^u \chi_2^d \psi_1^{0d} + \psi_2^u \chi_1^d \psi_2^{0d}\right) &
I_{14}^{R,d} = \left(\psi_1^u \chi_1^d \xi_2^d \psi_1^{0d} + \psi_2^u \chi_2^d \xi_1^d \psi_2^{0d}\right) \\
I_4^{R,d} = \left(\left(\psi_1^d\right)^2 \psi_2^u \psi_2^{0d} + \left(\psi_2^d\right)^2 \psi_1^u \psi_1^{0d}\right) &
I_{15}^{R,d} = \left(\psi_1^d \chi_1^u \xi_2^d \psi_1^{0d} + \psi_2^d \chi_2^u \xi_1^d \psi_2^{0d}\right)\\
I_5^{R,d} = \left(\psi_1^u \psi_2^{0d} +  \psi_2^u \psi_1^{0d}\right) \left(\psi_1^d \psi_2^d\right) &
I_{16}^{R,d} = \left(\psi_1^d \chi_1^d \xi_2^u \psi_1^{0d} + \psi_2^d \chi_2^d \xi_1^u \psi_2^{0d}\right) \\
I_6^{R,d} = \left(\psi_1^d \psi_2^{0d} \chi_1^d \chi_2^u + \psi_2^d \psi_1^{0d} \chi_2^d \chi_1^u\right) &
I_{17}^{R,d} = \eta^d \left(\chi_1^u \xi_1^d \psi_1^{0d} - \chi_2^u \xi_2^d \psi_2^{0d}\right) \\
I_7^{R,d} = \left(\psi_1^d \psi_2^{0d} \chi_2^d \chi_1^u + \psi_2^d \psi_1^{0d} \chi_1^d \chi_2^u\right) &
I_{18}^{R,d} = \eta^d \left(\chi_1^d \xi_1^u \psi_1^{0d} - \chi_2^d \xi_2^u \psi_2^{0d}\right) \\
I_8^{R,d} = \left(\psi_1^u \psi_2^{0d} + \psi_2^u \psi_1^{0d}\right) \left(\chi_1^d \chi_2^d\right) &
I_{19}^{R,d} = \eta^u \left(\chi_1^d \xi_1^d \psi_1^{0d} + \chi_2^d \xi_2^d \psi_2^{0d}\right) \\
I_9^{R,d} = \left(\psi_1^d \psi_2^{0d} \xi_1^d \xi_2^u + \psi_2^d \psi_1^{0d} \xi_2^d \xi_1^u\right) &
I_{20}^{R,d} = \eta^d \left(\xi_1^d \xi_1^u \psi_2^{0d} - \xi_2^d \xi_2^u \psi_1^{0d}\right) \\
I_{10}^{R,d} = \left(\psi_1^d \psi_2^{0d} \xi_2^d \xi_1^u + \psi_2^d \psi_1^{0d} \xi_1^d \xi_2^u\right) &
I_{21}^{R,d} = \eta^u \left(\left(\xi_1^d\right)^2 \psi_2^{0d} + \left(\xi_2^d\right)^2 \psi_1^{0d}\right) \\
I_{11}^{R,d} = \left(\psi_1^u \psi_2^{0d} + \psi_2^u \psi_1^{0d}\right) \left(\xi_1^d \xi_2^d\right) \; .&
\end{array}
\end{equation}
The second set reads
\begin{equation}
\begin{array}{ll}
I_1^{S,d} = \sigma \left(\psi_1^u \chi_1^d \varphi_2^{0d} + \psi_2^u \chi_2^d \varphi_1^{0d}\right) &
I_8^{S,d} = \left(\psi_1^u \chi_2^d \xi_1^d \varphi_2^{0d} + \psi_2^u \chi_1^d \xi_2^d \varphi_1^{0d}\right) \\
I_2^{S,d} = \sigma \left(\psi_1^d \chi_1^u \varphi_2^{0d} + \psi_2^d \chi_2^u \varphi_1^{0d}\right) &
I_9^{S,d} = \left(\psi_1^d \chi_2^u \xi_1^d \varphi_2^{0d} + \psi_2^d \chi_1^u \xi_2^d \varphi_1^{0d}\right) \\
I_3^{S,d} = \sigma \left(\psi_1^u \xi_2^d \varphi_1^{0d} + \psi_2^u \xi_1^d \varphi_2^{0d}\right) &
I_{10}^{S,d} = \left(\psi_1^d \chi_2^d \xi_1^u \varphi_2^{0d} + \psi_2^d \chi_1^d \xi_2^u \varphi_1^{0d}\right) \\
I_4^{S,d} = \sigma \left(\psi_1^d \xi_2^u \varphi_1^{0d} + \psi_2^d \xi_1^u \varphi_2^{0d}\right) &
I_{11}^{S,d} = \left(\left(\chi_1^d\right)^2 \psi_2^u \varphi_2^{0d} + \left(\chi_2^d\right)^2 \psi_1^u \varphi_1^{0d}\right) \\
I_5^{S,d} = \sigma \eta^d \left(\xi_1^u \varphi_1^{0d} -\xi_2^u \varphi_2^{0d}\right) &
I_{12}^{S,d} = \left(\chi_1^d \chi_1^u \psi_2^d \varphi_2^{0d} + \chi_2^d \chi_2^u \psi_1^d \varphi_1^{0d}\right) \\
I_6^{S,d} = \sigma \eta^u \left(\xi_1^d \varphi_1^{0d} +\xi_2^d \varphi_2^{0d}\right) &
I_{13}^{S,d} = \eta^d \left(\chi_1^d \chi_1^u \varphi_1^{0d} - \chi_2^d \chi_2^u \varphi_2^{0d} \right) \\
I_7^{S,d} = \left(\left(\psi_1^d\right)^2 \psi_1^u \varphi_2^{0d} + \left(\psi_2^d\right)^2 \psi_2^u \varphi_1^{0d}\right) &
I_{14}^{S,d} = \eta^u \left(\left(\chi_1^d\right)^2 \varphi_1^{0d} + \left(\chi_2^d\right)^2 \varphi_2^{0d} \right)
\end{array}
\end{equation}
and finally $I_k^{T,d}$ are given by
\begin{equation}
\begin{array}{ll}
I_1^{T,d} = \sigma \left(\psi_1^u \xi_1^d \rho_2^{0d} + \psi_2^u \xi_2^d \rho_1^{0d}\right) &
I_9^{T,d} = \left(\chi_1^u \xi_1^d \psi_2^d \rho_2^{0d} + \chi_2^u \xi_2^d \psi_1^d \rho_1^{0d}\right) \\
I_2^{T,d} = \sigma \left(\psi_1^d \xi_1^u \rho_2^{0d} + \psi_2^d \xi_2^u \rho_1^{0d}\right) &
I_{10}^{T,d} = \left(\chi_1^d \xi_1^u \psi_2^d \rho_2^{0d} + \chi_2^d \xi_2^u \psi_1^d \rho_1^{0d}\right) \\
I_3^{T,d} = \sigma \eta^u \left(\chi_1^d \rho_1^{0d} + \chi_2^d \rho_2^{0d}\right) &
I_{11}^{T,d} = \left(\chi_1^d \xi_1^d \psi_2^u \rho_2^{0d} + \chi_2^d \xi_2^d \psi_1^u \rho_1^{0d}\right) \\
I_4^{T,d} = \sigma \eta^d \left(\chi_1^u \rho_1^{0d} - \chi_2^u \rho_2^{0d }\right) &
I_{12}^{T,d} = \left(\left(\xi_1^d\right)^2 \psi_1^u \rho_1^{0d} + \left(\xi_2^d\right)^2 \psi_2^u \rho_2^{0d}\right) \\
I_5^{T,d} = \eta^d \left(\psi_1^d \psi_1^u \rho_1^{0d} - \psi_2^d \psi_2^u \rho_2^{0d}\right) &
I_{13}^{T,d} = \left(\xi_1^u \xi_1^d \psi_1^d \rho_1^{0d} + \xi_2^u \xi_2^d \psi_2^d \rho_2^{0d}\right) \\
I_6^{T,d} = \eta^u \left(\left(\psi_1^d\right)^2 \rho_1^{0d} + \left(\psi_2^d\right)^2 \rho_2^{0d}\right) &
I_{14}^{T,d} = \eta^d \left( \chi_2^u \xi_1^d \rho_1^{0d} - \chi_1^u \xi_2^d \rho_2^{0d}\right) \\
I_7^{T,d} = \left(\psi_1^u \left(\chi_1^d\right)^2 \rho_2^{0d} + \psi_2^u \left(\chi_2^d\right)^2 \rho_1^{0d}\right) &
I_{15}^{T,d} = \eta^d \left( \chi_2^d \xi_1^u \rho_1^{0d} - \chi_1^d \xi_2^u \rho_2^{0d}\right) \\
I_8^{T,d} = \left(\psi_1^d \chi_1^u \chi_1^d \rho_2^{0d} + \psi_2^d \chi_2^u \chi_2^d \rho_1^{0d}\right) &
I_{16}^{T,d} = \eta^u \left( \chi_2^d \xi_1^d \rho_1^{0d} + \chi_1^d \xi_2^d \rho_2^{0d}\right) \; .
\end{array}
\end{equation}
To actually calculate the shifts of the VEVs we
take the parameterization given in \Eqref{eq:VEVshifts} and plug this into the $F$-terms arising from 
the corrected superpotential. 
We then linearize the equations in $\delta \rm VEV$
and $1/\Lambda$ and can derive the following for
the shifts of the flavons with index $u$ from the $F$-terms of the driving fields $\psi^{0u}_{1,2}$,
$\varphi^{0u}_{1,2}$ and $\rho^{0u}_{1,2}$
\begin{align}
&b_u \left(v^u \delta w_2^u - w^u \delta v^u\right) + \frac{1}{\Lambda} \left\{
r_1^u  x^2 v^d    + 
r_2^u  x v^d w^d  +
(v^d)^3 \mathrm{e}^{-\frac{\pi i m}{7}} \left[r_3^u  - r_9^u \left(\frac{e_d z^d}{f_d w^d} \right)^2 \right] \right.\\ \nonumber &+
(v^u)^3 \left[r_4^u + r_{10}^u \left(\frac{e_u z^u}{f_u w^u} \right)^2 \right] +
r_5^u (w^d)^2 v^d +
r_6^u (w^u)^2 v^u +
v^d (z^d)^2 \left[r_7^u-r_{13}^u \frac{e_d}{f_d} -
r_{15}^u  \left( \frac{e_d z^d}{f_d w^d} \right)\right] \\ \nonumber &+ \left.
v^u (z^u)^2 \left[r_8^u-r_{14}^u \frac{e_u}{f_u} -
r_{16}^u  \left(\frac{e_u z^u}{f_u w^u}\right)\right]  +
r_{11}^u v^d w^d z^d +
r_{12}^u v^u w^u z^u   \right\} = 0
\end{align}
\begin{align}
&b_u \left(v^u \delta w_1^u + w^u \delta v^u\right) + 
\frac{1}{\Lambda} \left\{
r_1^u  \mathrm{e}^{-\frac{\pi i m}{7}} x^2 v^d    + 
r_2^u  \mathrm{e}^{-\frac{\pi i m}{7}} x v^d w^d   +
(v^d)^3 \mathrm{e}^{-\frac{2 \pi i m}{7}} \left[r_3^u  - r_9^u \left(\frac{e_d z^d}{f_d w^d} \right)^2 \right] 
\right. \\ \nonumber &+
(v^u)^3 \left[r_4^u + r_{10}^u \left(\frac{e_u z^u}{f_u w^u} \right)^2 \right] +
r_5^u \mathrm{e}^{-\frac{\pi i m}{7}} (w^d)^2 v^d +
r_6^u (w^u)^2 v^u +
v^d (z^d)^2 \mathrm{e}^{-\frac{\pi i m}{7}} \left[r_7^u  -r_{13}^u \frac{e_d}{f_d} - r_{15}^u  \left( \frac{e_d z^d}{f_d w^d}\right)\right] \\ \nonumber &+ \left.
v^u (z^u)^2 \left[r_8^u-r_{14}^u \frac{e_u}{f_u}  -
r_{16}^u \left( \frac{e_u z^u}{f_u w^u}\right) \right] +
r_{11}^u \mathrm{e}^{-\frac{\pi i m}{7}} v^d w^d z^d +
r_{12}^u v^u w^u z^u   \right\}=  0 
\end{align}
\begin{align}
&a_u \left(v^u \delta w_2^u+ w^u \delta v^u\right) +c_u v^u \delta z_2^u+d_u z^u \left[\delta \eta^u-
\left( \frac{e_u v^u}{f_u w^u} \right) \delta z_1^u\right] + \frac{1}{\Lambda} \left\{
\mathrm{e}^{\frac{\pi i m}{7}} s_1^u x v^d w^d \right.\\ \nonumber &+
\mathrm{e}^{\frac{\pi i m}{7}} x v^d z^d \left[s_2^u - s_3^u \left( \frac{e_d z^d}{f_d w^d}\right) \right] +
s_4^u (v^d)^3+
s_5^u (v^u)^3+
v^d w^d z^d \mathrm{e}^{\frac{\pi i m}{7}} \left(s_6^u- s_{10}^u  \frac{e_d}{f_d}\right)+
v^u w^u z^u \left(s_7^u- s_{11}^u \frac{e_u}{f_u}\right) \\ \nonumber & + \left.
\mathrm{e}^{\frac{\pi i m}{7}} s_8^u v^d (w^d)^2+
s_9^u v^u (w^u)^2 \right\}= 0 
\end{align}
\begin{align}
&a_u v^u \delta w_1^u+c_u \left(v^u \delta z_1^u+z^u \delta v^u\right) +d_u z^u 
\left[\delta \eta^u-\left( \frac{e_u v^u }{f_u w^u} \right)\delta z_2^u\right] +\frac{1}{\Lambda} \left\{ 
\mathrm{e}^{-\frac{2 \pi i m}{7}} s_1^u x v^d w^d \right.\\ \nonumber &+
\mathrm{e}^{-\frac{2 \pi i m}{7}}  x v^d z^d \left[s_2^u - s_3^u \left( \frac{e_d z^d}{f_d w^d}\right) \right]+
s_4^u \mathrm{e}^{-\frac{3 \pi i m}{7}} (v^d)^3 +
s_5^u (v^u)^3 +
v^d w^d z^d \mathrm{e}^{-\frac{2 \pi i m}{7}}  \left(s_6^u - s_{10}^u \frac{e_d}{f_d}\right) \\ \nonumber &+ \left.
v^u w^u z^u \left(s_7^u- s_{11}^u \frac{e_u}{f_u}\right) +
s_8^u \mathrm{e}^{-\frac{2 \pi i m}{7}} v^d (w^d)^2 +
s_9^u v^u (w^u)^2 \right\}
= 0
\end{align}
\begin{align}
&f_u w^u \delta \eta^u + e_u \left(v^u \delta z_2^u - \frac{v^u z^u}{w^u} \delta w_1^u+z^u \delta v^u\right) 
+\frac{1}{\Lambda}
\left\{
x v^d z^d \mathrm{e}^{\frac{2 \pi i m}{7}} \left( t_1^u - t_2^u \frac{e_d}{f_d} \right) \right.\\ \nonumber &-
t_3^u \mathrm{e}^{\frac{\pi i m}{7}} (v^d)^3 \left(\frac{e_d z^d}{f_d w^d} \right)-
t_4^u (v^u)^3 \left(\frac{e_u z^u}{f_u w^u} \right)+
t_5^u \mathrm{e}^{\frac{2 \pi i m}{7}} v^d (w^d)^2 +
t_6^u  v^u (w^u)^2 +
t_7^u \mathrm{e}^{\frac{2 \pi i m}{7}} v^d w^d z^d +
t_8^u  v^u w^u z^u \\ \nonumber & \left. -
v^d (z^d)^2 \mathrm{e}^{\frac{2 \pi i m}{7}} \left( t_9^u + t_{11}^u \frac{e_d}{f_d} \right) +
v^u (z^u)^2 \left( t_{10}^u - t_{12}^u \frac{e_u}{f_u} \right)\right\}
= 0 
\end{align}
\begin{align}
&f_u w^u \delta \eta^u + e_u v^u \left(\delta z_1^u - \frac{z^u}{w^u} \delta w_2^u \right) +\frac{1}{\Lambda} \left\{
x v^d z^d \mathrm{e}^{-\frac{3 \pi i m}{7}} \left( t_1^u - t_2^u \frac{e_d}{f_d} \right) \right.\\ \nonumber &+
t_3^u \mathrm{e}^{\frac{3 \pi i m}{7}} (v^d)^3 \left( \frac{e_d z^d}{f_d w^d} \right)-
t_4^u (v^u)^3 \left( \frac{e_u z^u}{f_u w^u}\right) +
t_5^u \mathrm{e}^{-\frac{3 \pi i m}{7}} v^d (w^d)^2 +
t_6^u  v^u (w^u)^2 +
t_7^u \mathrm{e}^{-\frac{3 \pi i m}{7}} v^d w^d z^d +
t_8^u  v^u w^u z^u \\ \nonumber & - \left.
v^d (z^d)^2 \mathrm{e}^{-\frac{3 \pi i m}{7}} \left( t_9^u + t_{11}^u \frac{e_d}{f_d} \right) +
v^u (z^u)^2 \left( t_{10}^u - t_{12}^u \frac{e_u}{f_u} \right)\right\}
= 0 
\end{align}
Note that we replaced the mass parameter $M^{u}_{\psi}$ by the VEV $w^u$.
Analogously, we replace the dimensionless coupling $m^{d}_{\psi}$
with the VEV $w^d$. We also frequently use the fact that $m$ is an odd integer in
order to simplify the phase factors appearing in the formulae.

Similarly, we can deduce another set of equations from the $F$-terms of the driving fields
$\psi^{0d}_{1,2}$, $\varphi^{0d}_{1,2}$ and $\rho^{0d}_{1,2}$ which gives rise to the shifts in
the VEVs of the flavons $\psi^{d}_{1,2}$, $\chi^{d}_{1,2}$, $\xi^{d}_{1,2}$, $\eta^d$ and $\sigma$
\begin{align}
&b_d \left(v^d \delta w_2^d - w^d \delta v^d\right) + \frac{1}{\Lambda} \left\{
r_1^d v^u x^2 +
r_2^d \mathrm{e}^{-\frac{\pi i m}{7}} v^d w^u x +
r_3^d \mathrm{e}^{\frac{\pi i m}{7}} v^u w^d x \right.\\ \nonumber &+
v^u (v^d)^2 \left[r_4^d +\mathrm{e}^{-\frac{\pi i m}{7}} r_5^d - r_{12}^d \mathrm{e}^{\frac{3 \pi i m}{7}} \left(\frac{e_d e_u z^d z^u}{f_d f_u w^d w^u} \right)- r_{13}^d \mathrm{e}^{-\frac{\pi i m}{7}} \left(\frac{e_d z^d}{f_d w^d}\right)^2 \right]+
v^d w^d w^u \left(\mathrm{e}^{\frac{\pi i m}{7}} r_6^d + \mathrm{e}^{-\frac{\pi i m}{7}} r_7^d \right) \\ \nonumber &+
r_8^d v^u (w^d)^2 +
v^d z^d z^u \left[\mathrm{e}^{\frac{2 \pi i m}{7}} r_9^d + \mathrm{e}^{-\frac{2 \pi i m}{7}} r_{10}^d - \mathrm{e}^{\frac{2 \pi i m}{7}} r_{18}^d \frac{e_d}{f_d} - \mathrm{e}^{-\frac{2 \pi i m}{7}} r_{20}^d \left(\frac{e_d z_d}{f_d w_d} \right)\right] \\ \nonumber &+ 
v^u (z^d)^2 \left[r_{11}^d + \mathrm{e}^{-\frac{3 \pi i m}{7}} r_{21}^d \left(\frac{e_u z^u}{f_u w^u} \right)\right] +
v^u w^d z^d \left[r_{14}^d \mathrm{e}^{\frac{\pi i m}{7}} - r_{19}^d \mathrm{e}^{-\frac{3 \pi i m}{7}} \left(\frac{e_u z^u}{f_u w^u} \right)\right] +
v^d w^u z^d \mathrm{e}^{\frac{\pi i m}{7}} \left[r_{15}^d \right.
\\ \nonumber &\left.\left. -r_{17}^d \left(\frac{e_d z^d}{f_d w^d} \right) \right]+
r_{16}^d \mathrm{e}^{-\frac{2 \pi i m}{7}} v^d w^d z^u \right\}
= 0
\end{align}
\begin{align}
&\mathrm{e}^{-\frac{\pi i m}{7}} b_d \left( v^d \delta w_1^d + w^d \delta v^d\right) + \frac{1}{\Lambda} \left\{ 
r_1^d v^u x^2 +
r_2^d v^d w^u x +
r_3^d \mathrm{e}^{-\frac{\pi i m}{7}} v^u w^d x \right.\\ \nonumber &+
v^u (v^d)^2 \left[r_4^d \mathrm{e}^{-\frac{2 \pi i m}{7}} +\mathrm{e}^{-\frac{\pi i m}{7}} r_5^d + r_{12}^d \mathrm{e}^{\frac{2 \pi i m}{7}} \left(\frac{e_d e_u z^d z^u}{f_d f_u w^d w^u} \right)- r_{13}^d \mathrm{e}^{-\frac{\pi i m}{7}} \left(\frac{e_d z^d}{f_d w^d}\right)^2 \right]+
v^d w^d w^u \left(\mathrm{e}^{-\frac{2 \pi i m}{7}} r_6^d + r_7^d \right) \\ \nonumber &+
r_8^d v^u (w^d)^2 +
v^d z^d z^u \left[\mathrm{e}^{-\frac{3 \pi i m}{7}} r_9^d + \mathrm{e}^{\frac{\pi i m}{7}} r_{10}^d - \mathrm{e}^{-\frac{3 \pi i m}{7}} r_{18}^d \frac{e_d}{f_d} - \mathrm{e}^{\frac{ \pi i m}{7}} r_{20}^d \left(\frac{e_d z_d}{f_d w_d} \right) \right]\\ \nonumber &+
v^u (z^d)^2 \left[r_{11}^d + \mathrm{e}^{\frac{3 \pi i m}{7}} r_{21}^d \left(\frac{e_u z^u}{f_u w^u} \right) \right]+
v^u w^d z^d \left[r_{14}^d \mathrm{e}^{-\frac{\pi i m}{7}} - r_{19}^d \mathrm{e}^{\frac{3 \pi i m}{7}} \left(\frac{e_u z^u}{f_u w^u} \right) \right]+
v^d w^u z^d \mathrm{e}^{-\frac{2 \pi i m}{7}} \left[ r_{15}^d\right.
\\ \nonumber & \left.\left. -r_{17}^d \left(\frac{e_d z^d}{f_d w^d} \right) \right]+
r_{16}^d \mathrm{e}^{\frac{ \pi i m}{7}} v^d w^d z^u \right\}
 = 0
\end{align}
\begin{align}
&\mathrm{e}^{\frac{\pi i m}{7}} \left[a_d \left(v^d \delta w_2^d+w^d \delta v^d\right)+c_d v^d \delta z_2^d - d_d z^d\left(\left[\frac{e_d v^d}{f_d w^d}\right] \delta z_1^d+ \delta \eta^d\right)\right]+\frac{1}{\Lambda} \left\{
s_1^d \mathrm{e}^{\frac{\pi i m}{7}} x v^u w^d\right. \\ \nonumber & + 
s_2^d x v^d w^u +
x v^u z^d \left[s_3^d \mathrm{e}^{\frac{2 \pi i m}{7}}-\mathrm{e}^{-\frac{2 \pi i m}{7}} s_6^d 
\left( \frac{e_u z^u}{f_u w^u} \right) \right] +
x v^d z^u \left[s_4^d \mathrm{e}^{-\frac{\pi i m}{7}}-s_5^d \mathrm{e}^{\frac{3 \pi i m}{7}} 
\left( \frac{e_d z^d}{f_d w^d} \right)\right] +
s_7^d (v^d)^2 v^u \\ \nonumber &+
s_8^d \mathrm{e}^{\frac{\pi i m}{7}} v^u w^d z^d +
v^d w^u z^d \mathrm{e}^{\frac{2 \pi i m}{7}} \left(s_9^d  -s_{13}^d \frac{e_d}{f_d} \right) +
s_{10}^d \mathrm{e}^{-\frac{\pi i m}{7}} v^d w^d z^u +
s_{11}^d \mathrm{e}^{\frac{2 \pi i m}{7}} v^u (w^d)^2 +
s_{12}^d v^d w^d w^u \\ \nonumber &- \left.
s_{14}^d \mathrm{e}^{-\frac{2 \pi i m}{7}} v^u (w^d)^2 \left(\frac{e_u z^u}{f_u w^u} \right)\right\}=0
\end{align}
\begin{align}
&\mathrm{e}^{-\frac{2 \pi i m}{7}} \left[a_d v^d \delta w_1^d+c_d \left(v^d \delta z_1^d+ z^d \delta v^d\right)-d_d z^d\left(\left[\frac{e_d v^d}{f_d w^d} \right]\delta z_2^d+ \delta \eta^d\right)\right] +\frac{1}{\Lambda} \left\{
s_1^d \mathrm{e}^{-\frac{\pi i m}{7}} x v^u w^d \right.\\ \nonumber & + 
s_2^d \mathrm{e}^{-\frac{\pi i m}{7}} x v^d w^u +
x v^u z^d \left[s_3^d \mathrm{e}^{-\frac{2 \pi i m}{7}}-\mathrm{e}^{\frac{2 \pi i m}{7}} s_6^d 
\left(\frac{e_u z^u}{f_u w^u} \right)\right] +
x v^d z^u \left[s_4^d +s_5^d \mathrm{e}^{\frac{3 \pi i m}{7}} \left(\frac{e_d z^d}{f_d w^d} \right)\right] 
\\ \nonumber &+ s_7^d \mathrm{e}^{-\frac{2 \pi i m}{7}} (v^d)^2 v^u +
s_8^d \mathrm{e}^{-\frac{\pi i m}{7}} v^u w^d z^d +
v^d w^u z^d \mathrm{e}^{-\frac{3 \pi i m}{7}} \left(s_9^d  -s_{13}^d \frac{e_d}{f_d} \right) +
s_{10}^d v^d w^d z^u +
s_{11}^d \mathrm{e}^{-\frac{2 \pi i m}{7}} v^u (w^d)^2 \\ \nonumber &\left. +
s_{12}^d \mathrm{e}^{-\frac{\pi i m}{7}} v^d w^d w^u -
s_{14}^d \mathrm{e}^{\frac{2 \pi i m}{7}} v^u (w^d)^2 \left( \frac{e_u z^u}{f_u w^u} \right)\right\}
=0
\end{align}
\begin{align}
&\mathrm{e}^{\frac{2 \pi i m}{7}} \left[e_d \left(v^d \delta z_2^d + z^d \delta v^d- \frac{v^d z^d}{w^d} \delta w_1^d\right)-f_d w^d \delta \eta^d\right] + \frac{1}{\Lambda} \left\{
t_1^d \mathrm{e} ^{\frac{2 \pi i m}{7}} x v^u z^d+
t_2^d x v^d z^u \right.\\ \nonumber &-
t_3^d \mathrm{e}^{-\frac{ \pi i m}{7}} x v^u z^u \left(\frac{e_u w_d}{f_u w_u}\right) -
t_4^d \mathrm{e}^{\frac{3 \pi i m}{7}} x v^d z^d \left(\frac{e_d w^u}{f_d w^d}\right) -
(v^d)^2 v^u \left[ t_5^d \mathrm{e}^{\frac{2 \pi i m}{7}} \left(\frac{e_d z^d}{f_d w^d} \right)
+ t_6^d \mathrm{e}^{-\frac{2 \pi i m}{7}} \left(\frac{e_u z^u}{f_u w^u} \right)\right] \\ \nonumber &+
t_7^d \mathrm{e}^{\frac{2 \pi i m}{7}} v^u (w^d)^2 +
t_8^d \mathrm{e}^{\frac{ \pi i m}{7}} v^d w^d w^u +
w^u z^d v^d \mathrm{e}^{\frac{ \pi i m}{7}} \left[t_9^d -t_{14}^d \left(\frac{e_d z^d}{f_d w^d}\right)\right] +
t_{10}^d v^d w^d z^u +
w^d z^d v^u \left[t_{11}^d \mathrm{e}^{\frac{3 \pi i m}{7}} \right.
\\ \nonumber &\left.\left.
- t_{16}^d \mathrm{e}^{-\frac{ \pi i m}{7}} 
\left(\frac{e_u z^u}{f_u w^u} \right) \right] -
t_{12}^d \mathrm{e}^{\frac{3 \pi i m}{7}} v^u (z^d)^2 +
z^u z^d v^d \mathrm{e}^{-\frac{3 \pi i m}{7}} \left(t_{13}^d + t_{15}^d \frac{e_d}{f_d} \right) \right\}
 = 0 
\end{align}
\begin{align}
&\mathrm{e}^{-\frac{3 \pi i m}{7}} \left[e_d \left(v^d \delta z_1^d-\frac{v^d z^d}{w^d} \delta w_2^d\right) 
- f_d w^d \delta \eta^d\right] + \frac{1}{\Lambda} \left\{
t_1^d \mathrm{e}^{-\frac{2 \pi i m}{7}} x v^u z^d+
t_2^d \mathrm{e}^{-\frac{\pi i m}{7}} x v^d z^u \right.\\ \nonumber &-
t_3^d \mathrm{e}^{\frac{ \pi i m}{7}} x v^u z^u \left( \frac{e_u w_d}{f_u w_u} \right)+
t_4^d \mathrm{e}^{\frac{3 \pi i m}{7}} x v^d z^d \left(\frac{e_d w^u}{f_d w^d} \right)+
(v^d)^2 v^u \left[ t_5^d \mathrm{e}^{\frac{3 \pi i m}{7}} \left(\frac{e_d z^d}{f_d w^d}\right) 
- t_6^d \left(\frac{e_u z^u}{f_u w^u} \right) \right]\\ \nonumber &+
t_7^d \mathrm{e}^{-\frac{2 \pi i m}{7}} v^u (w^d)^2 +
t_8^d \mathrm{e}^{-\frac{2 \pi i m}{7}} v^d w^d w^u +
w^u z^d v^d \mathrm{e}^{-\frac{2 \pi i m}{7}} \left[t_9^d -t_{14}^d \left(\frac{e_d z^d}{f_d w^d}\right)\right] +
t_{10}^d \mathrm{e} ^{-\frac{\pi i m}{7}} v^d w^d z^u \\ \nonumber &+\left.
w^d z^d v^u \left[t_{11}^d \mathrm{e}^{-\frac{3 \pi i m}{7}} - t_{16}^d \mathrm{e}^{\frac{\pi i m}{7}} 
\left(\frac{e_u z^u}{f_u w^u} \right)\right] -
t_{12}^d \mathrm{e}^{-\frac{3 \pi i m}{7}} v^u (z^d)^2 +
z^u z^d v^d \mathrm{e}^{\frac{2 \pi i m}{7}} \left(t_{13}^d + t_{15}^d \frac{e_d}{f_d} \right) \right\}
=0
\end{align}
One can infer the generic size of the shifts of the VEVs from these equations. In the case of no accidental
cancellation among the various terms present here we expect all of them to be of the order $\rm VEV^2/\La$ which
is $\epsilon \rm VEV \approx \epsilon^2 \La$ for all VEVs being of the order $\epsilon \La$ with 
$\epsilon \approx \lambda^2 \approx 0.04$.

\end{document}